\documentclass[preprint,5p,number,a4paper,sort&compress]{elsarticle}
\usepackage{graphicx}
\usepackage{epstopdf}
\usepackage{dcolumn}
\usepackage{amsmath}
\usepackage{mathptmx, courier, pifont}
\usepackage[scaled=0.92]{helvet}
\usepackage[T1]{fontenc}
\usepackage{textcomp}
\usepackage{color}
\usepackage{lineno,hyperref}
\modulolinenumbers[1]

\journal{nowhere, Condensed from the part of the thesis work of Zhuang Ge}

\begin{document}
\begin{frontmatter}
\title{Development of a large-area timing and position-sensitive foil-MCP detector for mass measurements at the Rare-RI Ring in RIKEN}
\cortext[mycorrespondingauthor]{Corresponding author}
\author[RIKEN,SAITAMA]
{Zhuang~Ge\corref{mycorrespondingauthor}\fnref{fn1}}
\fntext[fn1]{Present address: Department of Physics, University of Jyv\"askyl\"a, P.O. Box 35, FI-40014, Jyv\"askyl\"a, Finland\fnref{fn2}}
\fntext[fn2]{Condensed from the part of the thesis work of Zhuang Ge}
\ead{zhuang@ribf.riken.jp}
\ead{zhuang.z.ge@jyu.fi}
\ead{z.ge@gsi.de}
\address[RIKEN]{RIKEN Nishina Center, RIKEN, 2-1 Hirosawa, Wako, Saitama 351-0198, Japan}
\address[SAITAMA]{Department of Physics, Saitama University, Shimo-Okubo 255, Sakura-ku Saitama-shi, 338-8570,Japan}
\begin{abstract}
To achieve high precision and accuracy for mass measurements of exotic nuclei by Time-of-flight (TOF) methods: high-resolution beam-line magnetic-rigidity time-of-flight (B$\rho$-TOF) and in-ring isochronous mass spectrometry (IMS), a large-area electrostatic detector  which possesses high position resolution and good timing resolution at the same time  is developed at the Rare-RI Ring in RIBF, RIKEN Nishina Center, Japan.  Besides TOF mass measurements, the detector system will also be used for heavy ion beam trajectory monitoring or momentum measurements for both beam-line and in-ring at the Rare-RI Ring.
The  position and timing measurements of heavy ions are performed by detecting the secondary electrons (SEs) emitted from a conversion foil during the passage of the ion. The SEs are accelerated and bent with an angle of $90^{\circ}$ by electrostatic fields onto a micro-channel-plate (MCP) electron multiplier which is coupled with a position-sensitive delay-line anode.
The dependence of the timing and position resolution on applied high voltages of the detector potential plates  has been studied systematically via simulation and experimentally. An isochronous condition of secondary electron transmission in the electrostatic field of the detector is studied to optimize the structure of the detector for high performance. The best achieved timing resolution is better than 50 ps (in $\sigma$) and position resolution $\sim$ 1 mm (in $\sigma$) for 2 dimensions, respectively. The overall efficiency is $\sim$ 95$\%$ for heavy ion beam and  $\sim$ 75$\%$ for $\alpha$ particle from $^{241}$Am source.
\end{abstract}

\begin{keyword}
\texttt ~Isochronous mass spectrometry \sep B$\rho$-TOF \sep Storage ring \sep Micro-channel plates\sep Detector \sep Timing \sep Position-sensitive
\end{keyword}

\end{frontmatter}

\section{Introduction}
\label{Introduction}
Mass is a fundamental property of the nucleus, which results from the complex interplay of strong, weak and electro-magnetic interactions acting among all nucleons~\cite{Lunney:RMF}. New phenomena in nuclear physics such as shell structure, pairing correlations, decay and reaction properties have been discovered via nuclear mass measurements. Separation energies and reaction \textit{Q}-values can be deduced from atomic masses. The drip-lines (\textit{Sn} = 0 or \textit {Sp} = 0), which are defined as the borders of nuclear existence, are determined from the mass differences of neighboring nuclei. The final pathways of the nucleosynthesis, such as the r-process and rp-process in cosmos, are also governed by the masses and life-times of the nuclei involved. Strong interest in fast, high-accuracy and high-precision mass measurements for exotic nuclides due to their importance in nuclear astrophysics and nuclear structure studies, has triggered the development of a various of techniques for mass measurement around the world, such as: 
the Penning trap mass spectrometer ISOLTRAP/ISOLDE~\cite{Trap};
Schottky mass spectroscopy (SMS) or Isochronous mass spectroscopy (IMS) method at the storage rings ESR/GSI~\cite{BLT,Geissel} and CSRe/IMP~\cite{Xu}; 
Time-of-Flight (TOF) measurements implemented at several facilities including the SPEG spectrometer at GANIL~\cite{SPEG},
the TOFI spectrometer/LANL~\cite{TOFI}, S800 Spectrograph/NSCL~\cite{brho-tof};
Multi-reflection time-of-flight mass spectrometer (MR-TOF MS) at ISOLDE~\cite{MR-TOF}.

Foil-MCP detectors, by transporting induced SEs from a thin foil towards an MCP detector with different arrangements of the electromagnetic field, to deduce the timing or position information of the heavy ions, are widely utilized in mass measurement experiments.
Foil-MCP detectors for timing  have been longly used in mass measurement experiments of exotic nuclei in two heavy-ion storage ring facilities: the ESR/GSI~\cite{GSI-MCP} and CSRe/IMP~\cite{mei}, where precise mass measurements by revolution time measurement with electrostatic field and magnetic field crossly-arranged TOF detectors have been successfully performed in IMS.
Meanwhile, position-sensitive foil-MCP detector with parallel electrostatic field and magnetic field using for B$\rho$-TOF mass measurements at NSCL/MSU~\cite{matos} have also demonstrated its high performance and special characteristics for position measurements to deduce the momentum of exotic nuclei at a dispersive focal plane. All the above described detectors were mostly optimized for timing measurements at the cost of only timing information (ESR and CSRe) or only good position resolution  (NSCL), and relatively small detection active areas. In this paper, an electrostatic mirror-type foil-MCP detector with both high precision timing and good spatial resolution for the purpose of mass measurements and beam monitoring is developed. 
Similar detectors are used in many other laboratories around the world with many purposes and one of them is for mass identification studies, including fission fragments~\cite{e-mcp1,e-mcp2,e-mcp3}, decay of super heavy nuclides~\cite{e-mcp5} and elastic recoil~\cite{e-mcp6} due to its fast, compact and optimum timing resolution.

\section{Detector for new scheme of mass measurements}
\label{new scheme of mass measurements}
The Rare-RI Ring~\cite{R3}, dedicated for mass measurements in operation as an IMS at RIBF~\cite{yano}, has very recently taken into operation with projectile fragments and fission fragments at RIKEN in Japan. This new concept storage ring mass spectrometry is constructed based on a new scheme of IMS coupled with the individual injection method and velocity or B$\rho$ correction technique, which will achieve a mass accuracy on the order of 10$^{-6}$ within a measurement period of less than 1 ms. 

When the particle of interest with a mass-to-charge ratio of $\frac{m_{1}}{q_{1}}$  has the same momentum as that of a reference particle with a mass-to-charge ratio of $\frac{m_{0}}{q_{0}}$, the flight path length of these particles become identical in the isochronous storage ring. Consequently, the following equations hold for these particles~\cite{R3,ge2019}:
\begin{equation}
 \frac{m_{1}}{q_{1}} =  \frac{m_{0}}{q_{0}} \frac{T_{1}}{T_{0}}\sqrt{\frac{1-\beta_{1}^{2}}{{1-{(\frac{T_{1}}{T_{0}})^{2}\beta_{1}}}^{2}}}
 =  \frac{m_{0}}{q_{0}} \frac{T_{1}}{T_{0}}\sqrt{\frac{1-(\frac{T_{0}}{T_{1}})^{2}}{{{\left( \frac{\frac{m_{0}}{q_{0}}c}{(B\rho)_{0}}\right) }}^{2}}}.
\label{obit-IMS-vel}
\end{equation}
where $B\rho$ is the magnetic rigidity of the ion.
$\frac{m_{1}}{q_{1}}$ can be deduced from the ratio $\frac{m_{0}}{q_{0}}$  of the reference nucleus with the revolution time ($T_{0}$ for reference nucleus with an isochronous optics, $T_{1}$ for non-isochronous nucleus) and velocity ($\beta$)/momentum ($B\rho$) measurements of all ions.

Based on the technique of IMS, mass measurements have been performed successfully at ESR/GSI~\cite{BLT,Geissel} and CSRe/IMP~\cite{Xu}. The isochronous condition can only be fulfilled exactly for one species of ion. The revolution times of the other kind of ions still depend on their velocity or momentum. The non-isochronicity effect can be corrected with the velocity (uncertainty $\sim\, 10^{-4}$) or momentum (uncertainty $\sim\, 10^{-4}$) of each ion determined in addition to the revolution time (uncertainty $\sim\, 10^{-6}$). In IMS experiments at ESR, this is verified with the momentum acceptance limited by a pair of slits (B$\rho$-tagging) at a dispersive plane in FRS, which results in a significantly improved resolving power~\cite{Geissel}. However, the transmission efficiency of the rare exotic nuclides is reduced dramatically by this approach, limiting its application for nuclides far from  the valley of stability which have very tiny production rates.
This limitation can be overcome by the measurement of the velocity or momentum of each stored ion before injection into the ring or inside the ring. 
One method (double-TOF) to measure the revolution time with extra determination of the velocity is by using two TOF detectors installed at the straight section of the storage rings. The original idea was firstly proposed at GSI~\cite{DoubleTOF} and realized at CSRe in IMP~\cite{XingYM}. The Collector Ring (CR) at next generation facility FAIR~\cite{ILIMA} and the Spectrometer Ring (SRing) at HIAF~\cite{HIAF} will both be equipped with double TOF detectors~\cite{ZhangWei} at the straight sections. Another method (in-ring B$\rho$-TOF) is the measurement of the ion's position in the dispersive arc section of a storage ring by using a position-sensitive low-energy-loss foil MCP detector or non-destructive position-sensitive cavity doublet to determine the ion position event-by-event to realize the correction by magnetic rigidity measurement.
At the Rare-RI Ring a new method 
 is utilized: the total TOF of an ion inside the ring is measured by the TOF detectors at its injection as a start TOF (using foil-MCP detector with low energy loss) and at the extraction part as the stop TOF outside the ring. The velocity and momentum of the injected ions can be measured before its injection by the foil-MCP detector at a dispersive plane with low energy loss and small angular scattering of the ions to ensure the accuracy of velocity and momentum measurements.
The BigRIPS-OEDO-SHARAQ and the injection beam-line of the Rare-RI Ring is originally designed for the particle selection, high efficient transmission and particle identification of secondary ions from in-flight fission or projectile fragmentation reactions. The possibility of employing the BigRIPS-OEDO or BigRIPS-OEDO-SHARAQ beam-line for mass measurements via B$\rho$-TOF technique is highly motivated by its high resolution at a dispersive focus (76 mm/$\%$ or 147 mm/$\%$). A TOF measurements at two achromatic foci of the high-resolution beam-line with extra magnetic-rigidity measurements for TOF corrections to realize the B$\rho$-TOF method, makes it possible for mass measurements in one same experimental run to combine with the extended in-ring IMS mass measurements, while utilizing the momentum and timing information (stop TOF of B$\rho$-TOF and start TOF of IMS TOF) at one focal plane in two different approaches~\cite{ge2019}.

To design a detector to realize the scheme of the two combined high-precision mass measurement approaches (B$\rho$-TOF and IMS) and to employ it  as a beam position monitor for fast beam tuning with high efficiency, several conditions are to be considered: (a) Very good time resolution is essential (less than 100 ps), (b) Low energy loss and small angular scattering of the heavy ions, (c) Large active area to cover large beam size if located at a dispersive focal plane or in the ring, (d) Large detection efficiency, (d) No magnetic field to disturb the isochronous magnetic field when using in the ring. 

To satisfy the mentioned conditions above, an electrostatic mirror-type detector equipped with high performance Micro-channel-plates (MCPs) coupled with 2-dimensional position-sensitive delay-line anode is developed. The detector is characterized by detecting secondary electrons (SEs) (induced by the particles) with an deflection angle of $90^{\circ}$ towards an MCP surface by a pair of electrostatic mirrors which are tilted $45^{\circ}$ relative to the beam direction, to reconstruct the timing and position information of heavy ions.

In the following sections, the structure, principle of timing and position measurements of the electrostatic detector, calibration of an delay-line MCP detector, offline and online tests of the electrostatic detector will be presented.

\begin{figure}[!hbt]\centering
	\includegraphics[angle=0,width=8.5 cm]{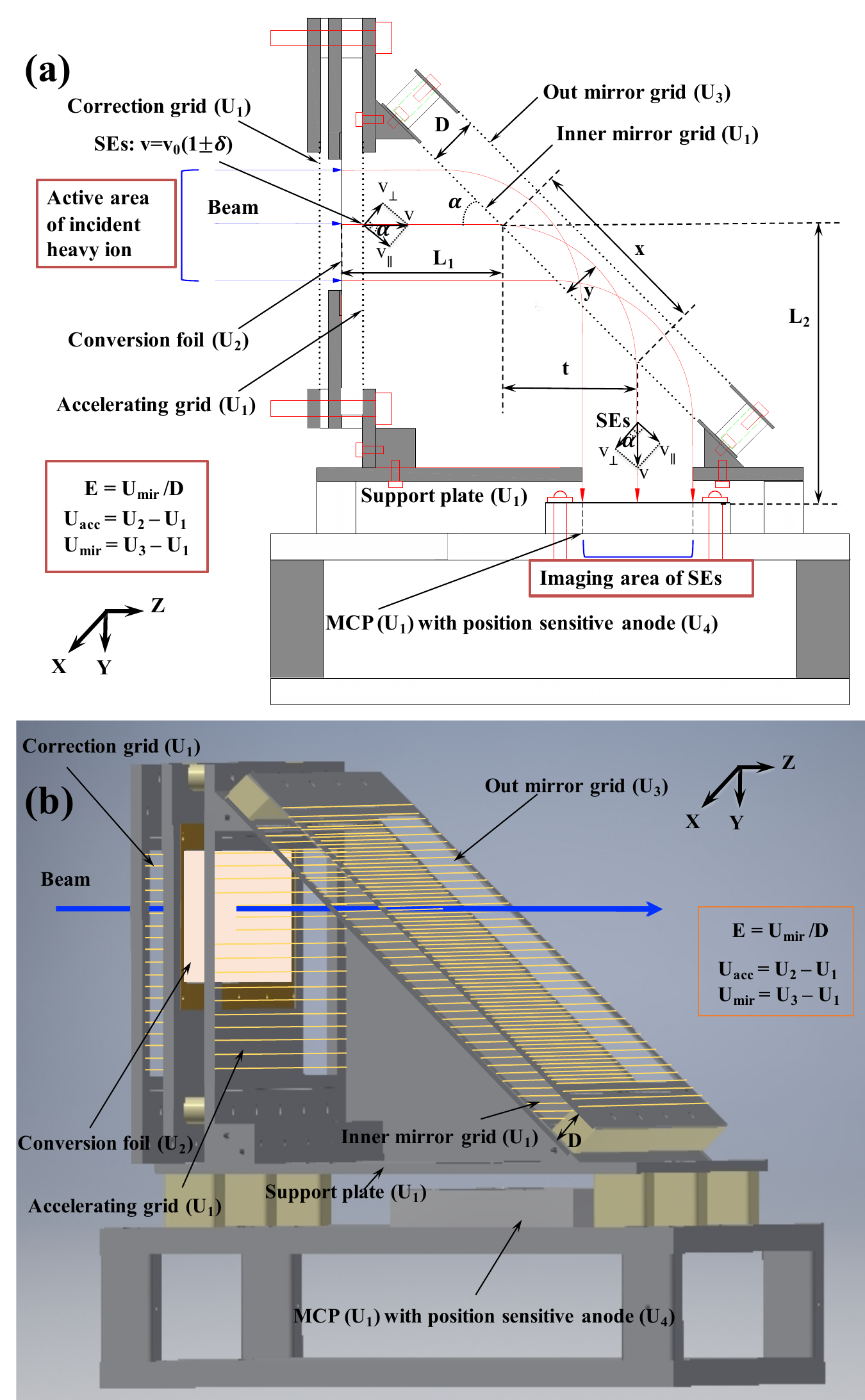}
	\caption{(Colour online) (a) Schematic view of the timing and position-sensitive detector with the electrostatic potential plates. (b) 3D CAD drawing of cross-sectional structure of electrostatic detector.
		\label{fig1}}
\end{figure}

\section{The electrostatic detector}
\label{The electrostatic detector}
In this section, the structure of the detector is described together with the motion of secondary electrons inside the detector driven by the homogeneous electrostatic field.
\subsection{Description of the electrostatic mirror detector}
\label{Description}
The operating principle of the detector with an electrostatic mirror is based on emission of SEs induced by the particles passing through the conversion foil of the detector. The detector system consists of a conversion foil, an accelerating grid, an electrostatic mirror, an equipotential house and a chevron-type MCP assembly as shown schematically in Fig.~\ref{fig1}.

%
When ions penetrate  the conversion foil, SEs are produced and then accelerated by the accelerating grid wires. After the acceleration,  SEs will enter the detector interior and pass a  field-free region and then are bent by the electrostatic mirror harp wires. Finally, the SEs are freely drifting to another field-free region and reach the MCP front surface and are detected by the coupled anode.  These processes are schematically shown in Fig.~\ref{fig1}, thus the detector can be divided into mainly three functional parts:
\begin{enumerate}
\item 
{SE generation and acceleration:}
Front wall with a conversion foil with a potential $U_2$, an accelerating plate with grids at the potential $U_1$,
\item 
{SE reflection:}
An electrostatic mirror plate at $U_3$  with  grids as back wall to reflect SEs,
\item 
{SE field-free region and SE detection:}
Bottom plate ($U_1$) with a circular or rectangular hole, two side walls ($U_1$) to keep equal potential inside and serve for fixing, and a chevron type MCP (front surface at $U_1$) with an delay-line anode ($U_4$) to detector SEs.
\end{enumerate}
Besides, an additional aluminium plate (at $U_1$) with grids, to balance the force of the foil and to prevent foil deformation or breakage due to the electrostatic force, is placed in front of the foil and have the same potential as the accelerating grid $U_1$. The construction materials of the detector  were selected with respect to both the high voltage requirements (dielectric strength) and vacuum conditions (material degassing) inside the detector. The conductive plates is made of aluminium and all the grids consist of gold-plated tungsten (W+Au) (40 $\mu$m in diameter). As the detector operates only with electrostatic filed to avoid the disturb of the isochronous filed in the storage ring, Rare-RI Ring, which is only consist of 24 dipoles~\cite{R3}. Carbon foil with 10 - 60 $\mu g/cm^{2}$ thicknesses or mylar foil coated with aluminium with thicknesses of 2 - 4 $\mu$m is used as the conversion foil for SEs emission. When using the detector for timing only, the MCP will be coupled with a timing anode. While if an two dimension position-sensitive delay-line anode is mounted below the output plate of MCP, it can be used for timing and position measurements at the same time.

\subsection{Electromagnetic motion of secondary electrons}
\label{motion}
To achieve high timing and spatial resolution for the electrostatic mirror detector, careful calculations of the relationships between the motion of SEs inside the detector and potentials supplied for the plates of the detector are required. 
In order to perform the calculations, the whole system, consisting of the conversion foil, accelerating grid, the mirror grids and the MCP detector is subdivided into three regions (see Fig.~\ref{fig1}). 
The first one extends from the accelerating grid to the inner mirror front side, which is a field free part. The second one is between the inner and out mirror surfaces, where a homogeneous electric field is assumed. The third region is from the inner mirror front side to the MCP front which is assumed to be field free too. Three particles, starting from the different points and in the same direction but with different energies are traced inside the system.
As magnitudes of potentials of several kV are applied to potential plates or grids to accelerate or bend the SEs, the whole system can be described non-relativistically. The typical initial energy of SEs is several eV (most of them) and in this section we neglected the initial energy of SEs for calculation.
After acceleration from the foil to the accelerating grid as depicted in up panel of Fig.~\ref{fig1}(a), SEs gain a velocity of $v_0$. At a certain time  when the velocity component parallel to the mirror of the SE is decelerated to zero and the electron reaches the maximal depth y inside the mirrors measured from the entrance mirror grid plane holds:
\begin{equation}
y=\cos{\alpha}^2 D\frac{U_{acc }}{U_{mir}}.\label{ymax},
\end{equation}
where $U_{acc}$ and $U_{mir}$ are, respectively the acceleration potentials between the conversion foil and the accelerating grid, and the mirror deflection potential between the mirrors, D the distance of the mirrors, $\alpha$ the tilted angle of the mirrors.
When escaping the mirror, the electron gains velocity and continues freely drifting towards the MCP detector through the field free region. As demonstrated in Fig.~\ref{fig1}(a), $x$, $y$ are parallel and vertical path length of SEs in the mirror parts, and t is the shift of SEs along the Z direction inside the mirror, satisfying the equations:
\begin{equation}
x = 2D\sin( {2\alpha })\frac{U_{acc}}{U_{mir}},
\end{equation}
\begin{equation}
t = x\sin {\alpha } = 2D\sin {\alpha }\sin ({2\alpha } )\frac{U_{acc}}{U_{mir}}.
\end{equation} 
For the convenience of calculation and design, the $\alpha$ is set to be $45^{\circ}$. When substituted for $\alpha$ with $45^{\circ}$, the following equations for $x$, $y$ and $t$ are obtained:
\begin{equation}\label{eq_motion}
x=2{D}\frac{U_{acc}}{U_{mir}},  \ 
y =\frac{{D}}{2}\frac{U_{acc}}{U_{mir}},  \ 
t = \sqrt{2}{D}\frac{U_{acc}}{U_{mir}}.  
\end{equation}
In principle, from the Eq~\eqref{eq_motion}and the condition $D \ge y$, it is obvious that the electrostatic mirror will reflect the SEs to the MCP front only if the following condition is satisfied:
\begin{equation}
\frac{U_{mir}}{U_{acc}}\geq0.5.\label{transparency}
\end{equation}

From the above analytical calculation of the  motion of SEs inside the detector, imaging position of the SEs induced by heavy ions from the foil onto the MCP front surface and total TOF inside the detector can be determined by the acceleration potentials $U_{acc}$ and the mirror deflection potential $U_{mir}$ added to the detector potential plates. Therefore, the position and timing information of the heavy ions when bombarding the foil can be reproduced by detecting the position and timing of SEs at the MCP detector.
\subsubsection{Isochronous condition}
\label{Isochronous condition}
To achieve high timing resolution, an isochronous condition~\cite{e-mcp2,APR,Ge2019} is considered. The total time of flight  of 3 parts: the foil to the inner mirror wires, bending path between the inner and outer mirror wires, free drift region from the inner mirror wires to the MCP surface. 
The total TOF of the probed particles (initial velocity of SEs considered to be ${v_0}\delta$) after exact solution of the equations of motion is given as:
\begin{equation}
\begin{aligned}
T=
T_0+  \left( \mp {\frac{L_1+L_2-\sqrt2D\frac{U_{mir}}{U_{acc}} }{v_0} \pm \frac{2m_e v_0 D\sin{\alpha}}{e\Delta{V}}} \right) \delta,
\label{eq_i}
\end{aligned}
\end{equation}
where $L_1$ and $L_2$, as shown in Fig.~\ref{fig1}, correspond to length from the conversion foil to the incident point of SEs at the inner mirror plate along  Z direction, the length between the incident point of SEs at the inner mirror plate with the MCP front surface along Y direction. 
To let the second term of the right part of the Eq.~(\eqref{eq_i}) to be zero, we get the isochronous condition in which the total TOF has no dependence of initial velocity variation of SEs:
 $D/(L_1+L_2)=0.236\frac{U_{mir}}{U_{acc}}$. This relation enables us to optimize the foil acceleration and mirror deflection potentials such that the SEs reach the MCP isochronously.

\section{Simulation}
\label{simulation}
To optimize the design and performance of the electrostatic mirror detector, a simulation of the transport of SEs induced from the conversion foil is done by SIMION software package~\cite{SIMION}. 
In this simulation procedure, energies of SEs were assigned to a range of 0-20 eV with a mean value $\sim$ 2 eV which corresponds to the main electrons emitted from the foil with a percentage of 85 \%~\cite{electron,electron2}. The angles lie between -90 and +90 degrees relative to the foil surface and the SEs mainly emitting from forward side of the foil are guided to the surface of MCP. Each parameter of the uniformly distributed electrons is generated using a random generator in SIMION. 

\begin{figure}[!hbt]\centering
	\includegraphics[angle=0,width=8.5 cm]{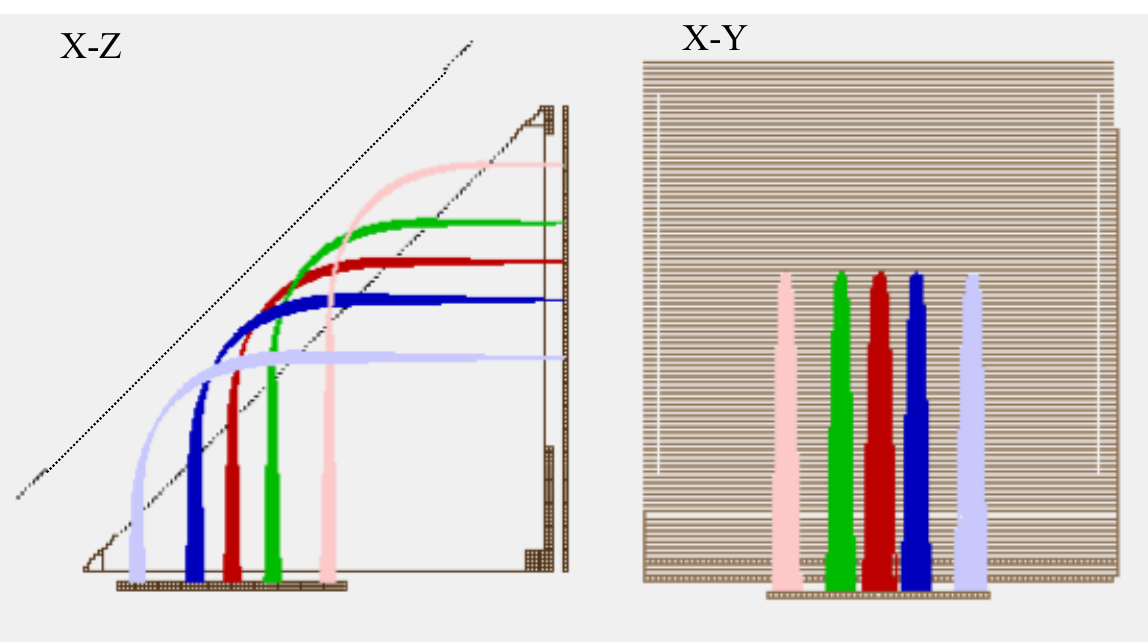}
	\caption{(Colour online) 5 points imaging of SEs from the foil onto the MCP surface  in the X-Z (left panel) and X-Y (right panel) view in the simulation.
		\label{fig2}}
\end{figure}

The real electric field usually contains disturbances compared to the perfectly designed electric field. Because of that, the timing resolution and spatial focusing can be different from that of the calculated ideal case. However, the analytical solution (simulation) can be useful to define the direction of the investigations.
In order to determine the timing and position resolution, SEs are assigned to groups and start from the same point of the foil. 5 points imaging of grouped SEs from the foil onto the MCP surface in the X-Z (left panel of Fig.\ref{fig2}) and X-Y (right panel  of Fig.\ref{fig2}) view in the simulation.
Different high voltage (HV) supplies are added to the potential plates or grids and varied during the simulation for different settings. The TOF distribution, and 2-dimension position distributions of initial SEs for each group from the foil reflected onto the MCP are fit by a Gaussian function. The peak width is characterized by the Full Width at Half of Maximum (FWHM) and it is defined as 2.355$\sigma$, where $\sigma$ is the uncertainty parameter of Gaussian distribution.

\begin{figure}[!hbt]\centering
	\includegraphics[angle=0,width=6.2 cm]{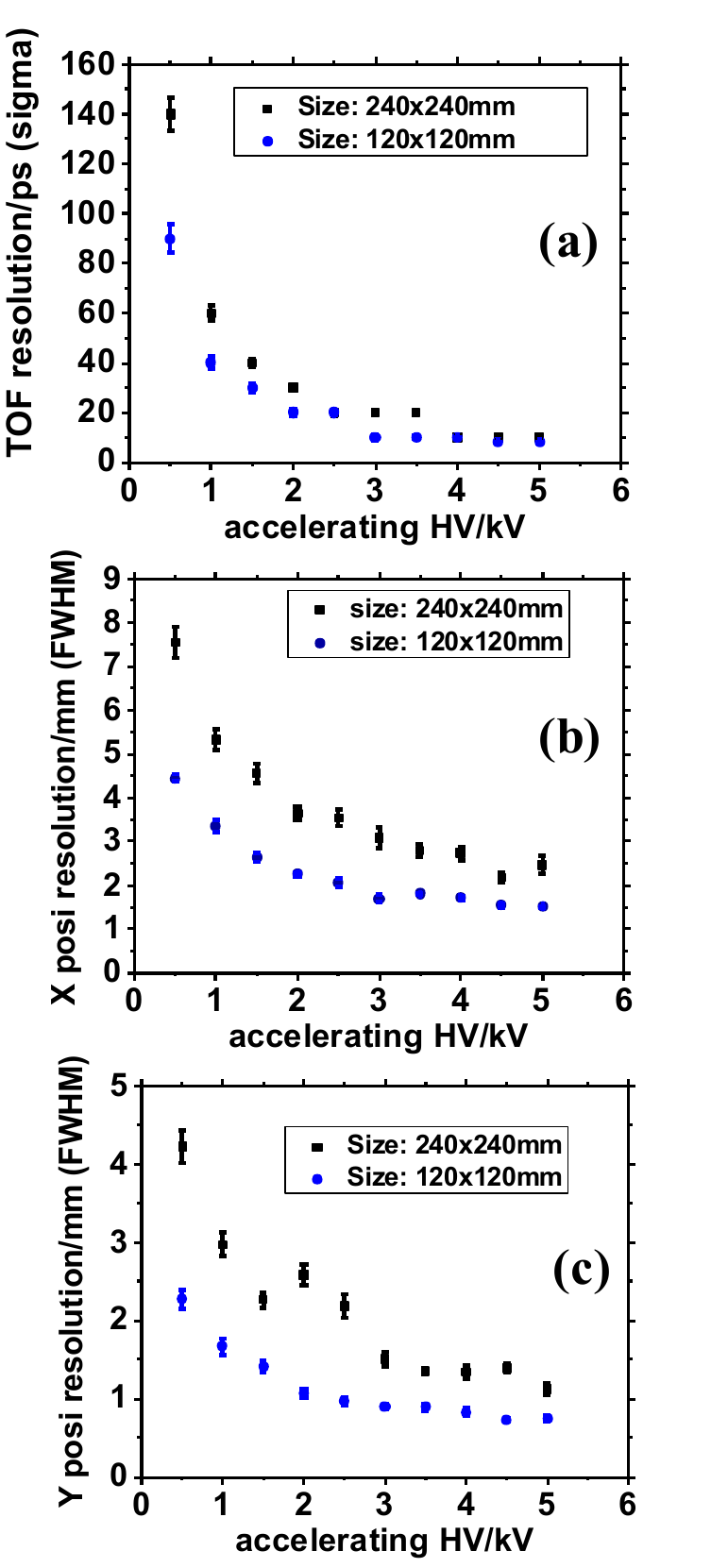}
	\caption{(Colour online) 
Comparison of timing resolution (a), X-coordinate position (b) and Y-coordinate position (c)  resolutions for detectors with different dimensions (120 mm $\times$ 120 mm and 240 mm  $\times$ 240 mm for the triangular structure) from simulation. The HV settings of different plates are all the same and the accelerating HV values are all negative in the simulation.
 \label{fig3}}
\end{figure}

Comparison of timing and position resolutions of different dimensions (120 mm $\times$ 120 mm and 240 mm $\times$ 240 mm for the triangular structure) are simulated and the results are demonstrated in Fig.~\ref{fig3}. It is obvious that the position resolutions and timing resolution are better for a smaller size of the detector, with the same settings for the HV supplies. As the total TOF and path length of SEs are smaller in a more compact structure, a smaller influence on the position resolutions and timing resolution will come from initial energy and angular distribution of SEs induced from the foil. The typical trends of timing and position resolutions for the mirror detector as shown Fig.~\ref{fig3} indicates that as the increasing of the accelerating HV, the timing and position resolutions will get improved accordingly and finally get nearly saturated at a plateau.

 \section{Calibration of a MCP detector with helical delay-lines}
\label{calibration}
\begin{figure}[!hbt]\centering
	\includegraphics[angle=0,width=9.5 cm]{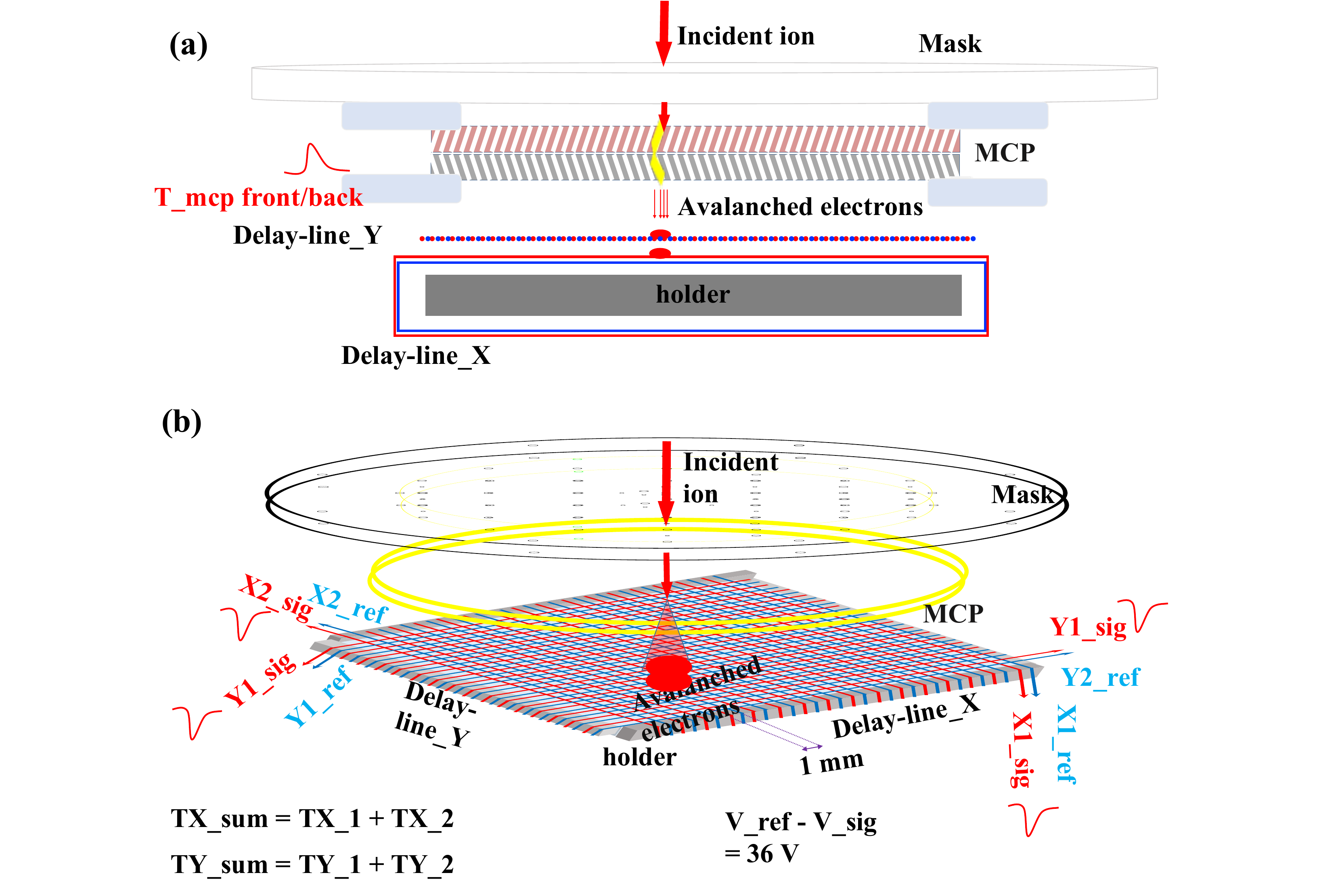}
	\caption{(Colour online) (a) shows schematic cross-sectional view of the setup for the calibration of the DLD system. (b) indicates the 3D imaging principle of the calibration setup.
			\label{fig4}}
\end{figure}

 %
The  helical delay-line anode of the delay-line MCP detector (DLD)~\cite{roentdek}, as shown in Fig.~\ref{fig4}, consists of a holder and two coils for X, Y directions.  Each coil (dual delay-line) with two wires is convoluted in parallel with a pitch of  1 mm around the holder, and one of the wires acts as a reference wire (Reference) and the other one as charge collection wire (Signal). The holder has a metal core with four checkered ceramic insulators at the edges. Around these insulators the delay-lines are convoluted, the first one in a direction with a smaller circumference (X), the second in the direction perpendicular to the first one with a larger circumference (Y). Signal is on a more positive potential (+36 V) than Reference. Thus the Signal wire is more attractive to the electrons coming from the MCP back plate and produces a larger signal. If a electron cloud propagates along a collecting wire, a current will  only be induced in the neighboring Reference wire and thereby can be registered. The  anode  holder acts as a reflecting plate, so that nearly all electrons from the charge cloud can be collected by the Signal wire.  The single pitch propagation time (for 1 mm) on the delay line is $\sim$ 1.24 ns for the DLD, thus the correspondence between 1 mm position distance and relative time delay in the 2-dimension image is twice of this value:  $\sim$ 2.48 ns. Reversely, relative time delay of 1 ns in the 2-dimensional image corresponds to $\sim$ 0.4  mm position distance.
Each line has a length of  approximately 120 mm (300 ns) and a resistance of $\sim$ 24 $\Omega$. 
\begin{table}\footnotesize
\caption{Typical high voltage settings for delay-line MCP detector. }
\centering
\begin{tabular}{ccc}
\hline
&Ion Detection mode & Electron Detection mode\\
\hline
MCP front &-2400 V& 0 V\\
MCP back & 0 V  &+2400 V\\
Anode Holder  &  0 V to +250 V & +2400 V to +2650 V\\
 Reference wires & +250 V & +2650 V\\
Signal wires & +286 V & +2686 V\\
\hline
\end{tabular}
\label{tab:HV-setting}
\end{table}


The front side and back side of the MCP assembly, delay-line anode wires and the holder are biased through the  feedthrough decouple (FT12)~\cite{roentdek} which connects the MCP front, back, anode wires, and holder with copper cables inside the chamber. The bias voltages applied to the front and back side of the MCP, the holder electrode and the delay-line anode wires are listed in Table~\ref{tab:HV-setting}. 
Between the front and back of the MCP, a bias voltage of 2400 V is applied, and the overall gain of the two MCPs is $\sim10^{7}$. The avalanched  electrons leaving the MCPs induces a fast positive signal on the back/front side of the MCPs and are collected by the signal wires of the X and Y delay lines as shown in Fig.~\ref{fig4}.  The reference wires wound next to the signal wires are biased with slightly different voltages (+36V) by a BA3 module from RoentDek~\cite{roentdek} and are used to suppress the electromagnetic noise picked up in the vacuum chamber. Five signals are read out: one from the MCP back or front side, two from the ends of the X delay line, and two from the ends of the Y delay line. The MCP back-side signal is inverted and then amplified twice by an photomultiplier amplifier (PM-AMP, Kaizu KN2104~\cite{kaizu}). The four anode signals are amplified twice by the same PM-AMP.
The gains of all channels after these amplifiers are $\sim$ 25 - 100. 
When carrying out the calibration for the DLD system, the source of ions used is usually the vacuum gauge with  low energy ions (close to 100 eV) produced or the  $^{241}$Am $\alpha$ sources put above the mask on the chamber when supplying the HV for the DLD system in the so-called ion mode as shown in Table~\ref{tab:HV-setting}. For the electron mode, as  ions from the gauge can scarcely arrive at the MCP front surface with applied voltage of 0 volt, the source applied is the $^{241}$Am $\alpha$ sources.  

\begin{figure}[!hbt]\centering
	\includegraphics[angle=0,width=6.0 cm]{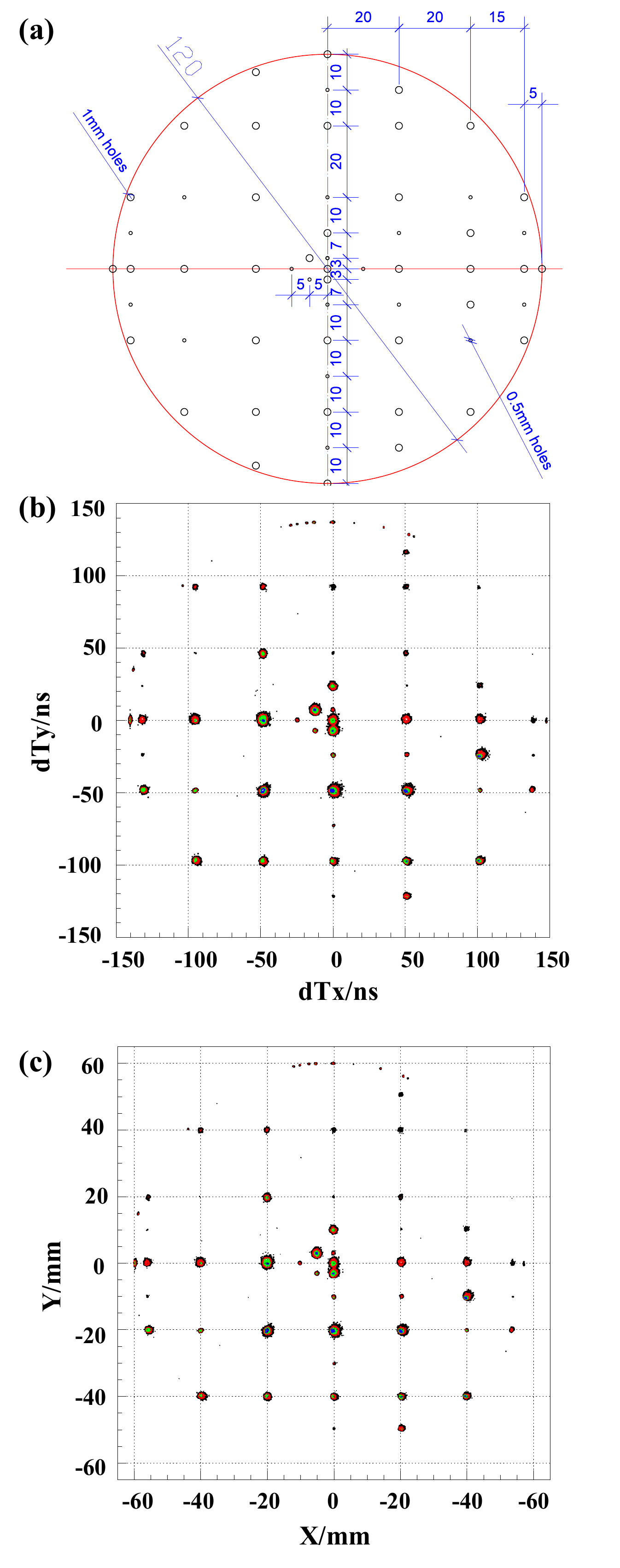}
	\caption{(Colour online) (a) The CAD drawing of effective area with $\phi$0.5 mm or $\phi$1 mm holes on the mask for calibration. (b) The 2-dimentional spectrum of raw signal imaging by time difference of the x- and y-directions with contour display. (c) The calibrated  2-dimentionial position imaging spectrum of the collimated ions through the mask with contour display.
			\label{fig5}}
\end{figure}



The position where the incoming particle hits the MCP detector is determined by the time difference between the two signals from each end of the same delay-line anode. 
The MCP position readout (X, Y) can be expressed as:
\begin{equation}
X =a_{x} (T_{X1} - T_{X2}) + b_{x},\,
Y= a_{y} (T_{Y1} - T_{Y2}) + b_{y},
\label{ecoding-x-y-delayline}
\end{equation}
where $T_{X1}, T_{X2}, T_{Y1}, T_{Y1}$ are the timings of the four delay-line anode signals in ns. The ns-to-mm conversion factor $a_{x}, a_{y}$ can be calibrated with a collimated mask, and timing offsets $b_{x}, b_{y}$, which are introduced in the pulse propagation and amplification, can also be deduced at the same time. 
Although this basic correction scheme by Eq.~(\ref{ecoding-x-y-delayline}) can not restore rotations and non-linear distortions of the MCP image, it makes the calibration algorithm easier to implement. A typical setup for the calibration is shown in Fig.~\ref{fig4}.

As the total length of the delay-lines are fixed, the sum of the propagation times from the charge impact position to the two ends of the delay-lines are fixed, independent of  position at which the event happens. Thus, the timing sums:
\begin{equation}
T_{Xsum} =  T_{X1} +T_{X2}, \
T_{Ysum} =  T_{Y1} +T_{Y2}
\label{calibration_delay_line} 
\end{equation}
are both expected to be constants, where  $T_{X1}$, $T_{X2}$, $T_{Y1}$, $T_{Y2}$  are the timing signals from both ends  of the 2-dimensional delay-lines relative to the MCP back-side signal, which is used as trigger. 
When conducting experiment with  position  measurement or  calibration of the DLD, the sum condition on the timing of the signals is imposed for event selection in order to rule out false triggers by electronic noise and pile-up events. 

\begin{figure}[!hbt]\centering
	\includegraphics[angle=0,width=0.4\textwidth]{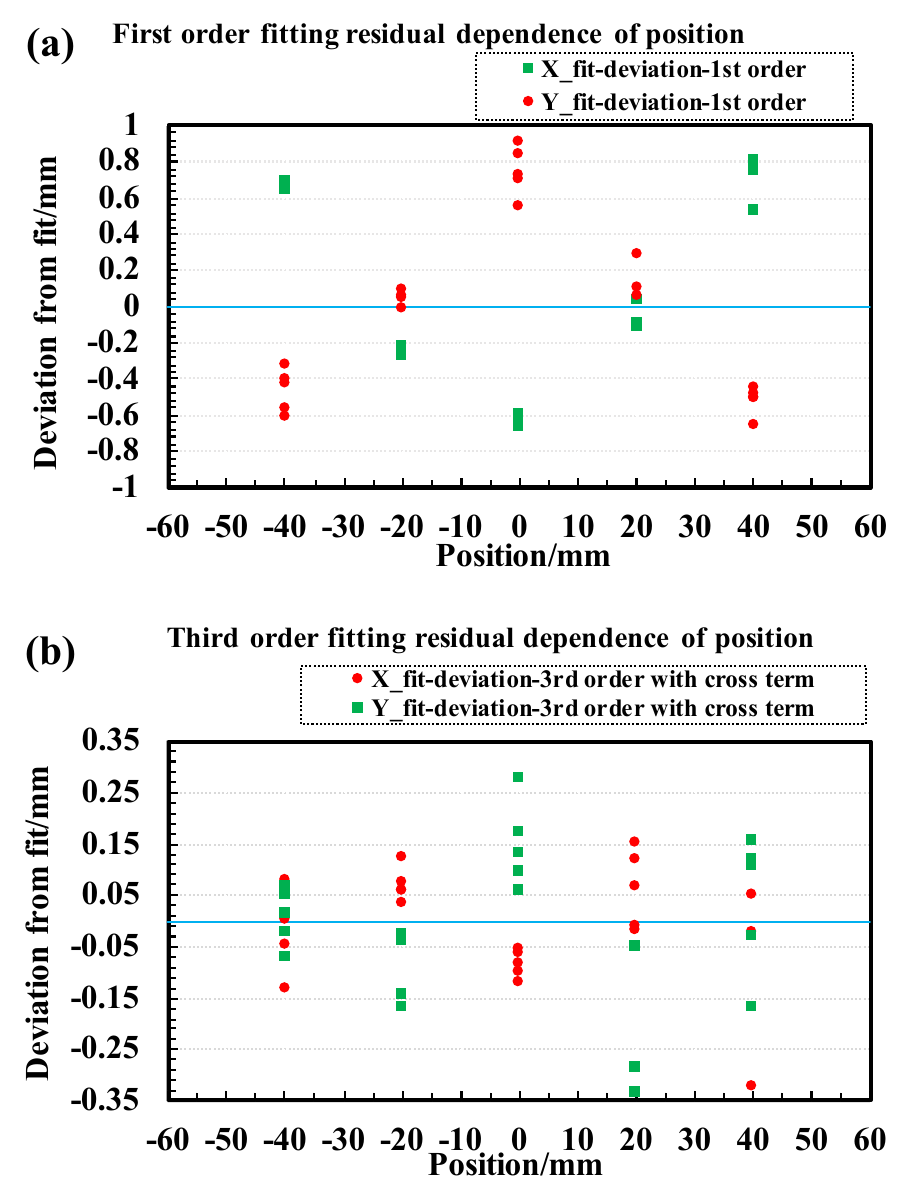}
	\caption{(Colour online) Fitting residual dependence of position for first order calibration (a) and third order calibration with cross terms (b) in fitting.
	\label{fig6}}
\end{figure}

The physical positions ($X_{p}$, $Y_{p}$) of the hole points along the delay-lines on the mask are determined from the designed values as shown in Fig.~\ref{fig5}(a). A first order calibration function similar to  Eq.~(\ref{ecoding-x-y-delayline}) is constructed to correct the measured timing difference of each center positions ($dT_{X}$, $dT_{Y}$) on the holes to their physical positions. The parameters $ a_{x},\,b_{x},\ a_{y},\,b_{y}$ from fitting are then applied to Eq.~(\ref{ecoding-x-y-delayline}) to calculate the calibration points with measured timing difference from the both ends of each delay-lines.

To estimate the distortions of the MCP image, we calculate the first order fitting deviations between the coordinates of the calibration points on the MCP image and their physical positions as shown at  Fig.~\ref{fig6}(a).
The mean absolute deviation is 0.44 mm and 0.42 mm for X- and Y- directions,  and the maximum deviation is 0.85  mm and 0.92 mm for X- and Y- directions without large overall rotation of the mask when observing the whole MCP image. 
To achieve a much more precise and accurate calibration for the DLD system, for each hole, we use a two-dimension  polynomials with 10 parameters  to transform the high-accuracy timing difference of each positions ($dT_{X},dT_{Y}$) of the calibration points to their physical positions ($X_{p},Y_{p}$) as close as possible. 
The correction functions are:
\begin{equation}
\begin{split}
X_{c} = p_{0} + p_{1}dT_{X} +p_{2}dT_{y}+p_{3}dT_{X}^{2}+p_{4}dT_{X}dT_{Y}\\+p_{5}dT_{Y}^{2}+p_{6}dT_{X}^{3}+p_{7}dT_{Y}^{3}+p_{8}dT_{X}^{2}dT_{Y}+p_{9}dT_{X}dT_{Y}^{2}
\end{split}
\label{calibration_x-higherorder} 
\end{equation}
and
\begin{equation}
\begin{split}
Y_{c} = q_{0} + q_{1}dT_{X} +q_{2}dT_{y}+q_{3}dT_{X}^{2}+q_{4}dT_{X}dT_{Y}\\+q_{5}dT_{Y}^{2}+q_{6}dT_{X}^{3}+q_{7}dT_{Y}^{3}+q_{8}dT_{X}^{2}dT_{Y}+q_{9}dT_{X}dT_{Y}^{2},
\end{split}
\label{calibration_y-higherorder} 
\end{equation}
where $dT_{X}=T_{X1} -T_{X2}$ and $dT_{Y}=T_{Y1} -T_{Y2}$ are timing difference of the both delay-line signals of X and Y layers.
In Eq.~(\ref{calibration_x-higherorder}) and ~\ref{calibration_y-higherorder}, ($X_{c},Y_{c}$) is the corrected position of a calibration point with non-corrected timing difference ($dT_{X},dT_{Y}$), and $ p_{i}$ and $q_{i} $ ( i= 0-10) are parameters determined by a fit using all the calibration points on an effective region ($-45\,mm\le X \le 45\,mm, \, -45\,mm\le Y \le 45\,mm$) of the mask and setting ($X_{c}, Y_{c}$) to their physical positions. The regions near the outer edge of the MCP which have not good linearity are ruled out from the calibration analysis. 

\begin{figure}[!htp]
\centering
	\includegraphics[angle=0,width=8.0 cm]{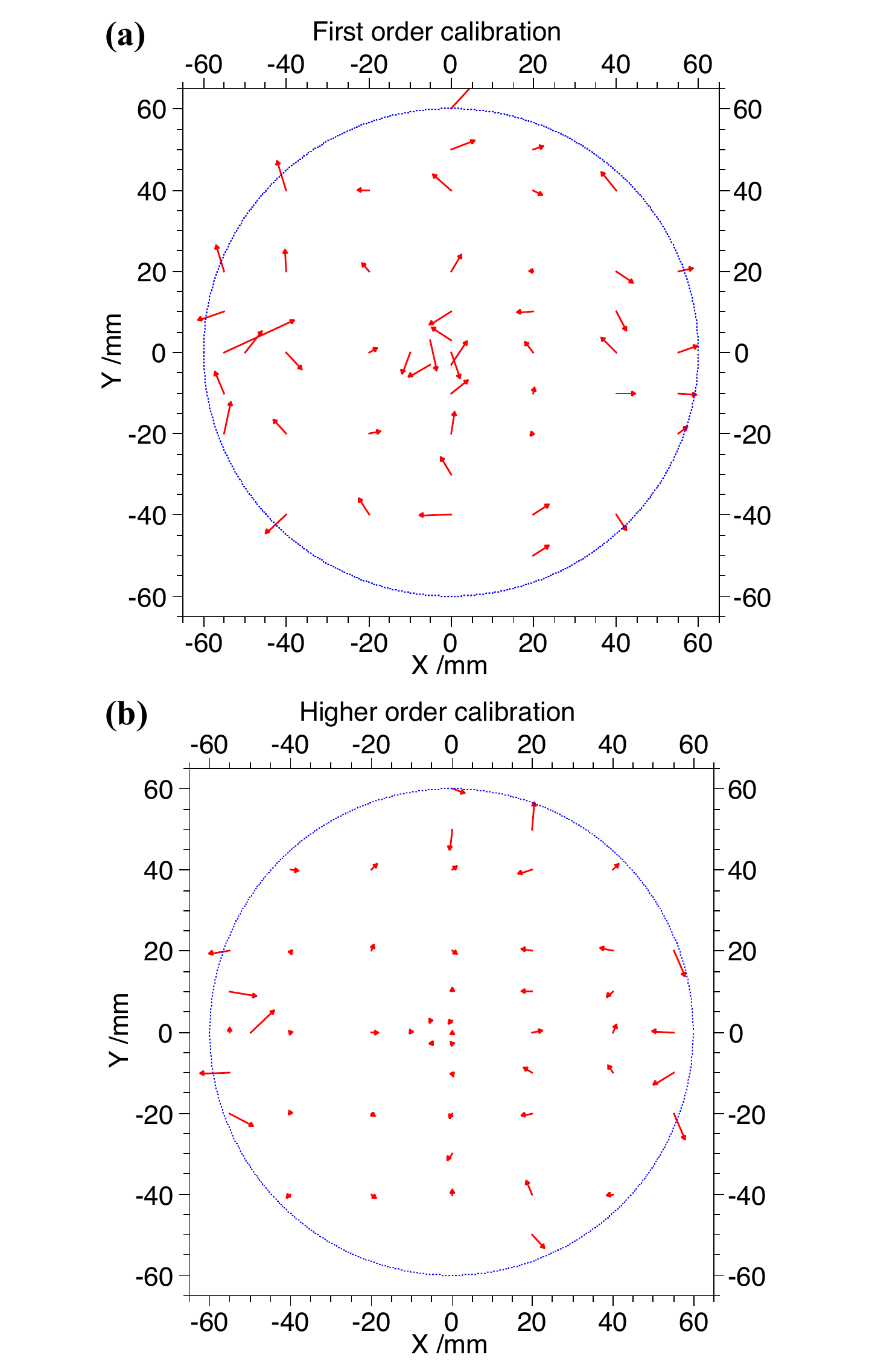}
	\caption{Vector field map of the correction vector $\vec {V_{ij}}$ created from the difference of the expected and the calibrated/measured mean values for each hole center spot. (a) shows a first order correction map and (b) illustrates a second order correction map. The lengths (magnitude) of the vectors are enlarged by a factor of 5 for  better visibility. The edge of the active area of the DLD is marked by the dashed blue circle.
		\label{fig7}}
\end{figure}

To test the performance of this position calibration method, we utilize the calibration parameters from one calibration run to correct another data set which was taken under the same conditions but statistically independent from the calibration run.  
The deviations of x and y coordinates from their physical values were calculated. The distributions of the deviations with the uncertainties of the measurements in the X and Y directions are shown in Fig.~\ref{fig6}(b). The average absolute deviations (accuracy) of a second order or third order correction with/without cross terms amounts to $\le$ 82 $\mu\, m$ and $\le$ 147 $\mu\, m$ for x and y coordinates which are much smaller than those from the first other  correction with values of $\sim$ 0.4 mm. The root-mean-square (RMS, resolution: $\sigma_{x}$, $\sigma_{y}$) of the higher order deviation distributions are  all $\le$ 0.068 mm and 0.142 mm for both x and y coordinates, and in this test the maximum deviation is < 0.4 mm as shown in Fig.~\ref{fig7}. For the radial position resolution, we transform  X- and  Y-resolution as follows:
\begin{equation}
R_{xy} = \sqrt{{\sigma_{x}}^{2} + {\sigma_{y}}^{2}}.
\label{deltaxy-res}
\end{equation}
Therefore, the radial position resolution: $R_{xy}$ = $\sqrt{0.68^{2} + 0.142^{2}}$ = 0.157 mm.
 
From the  comparison of first order and higher order calibration method, the higher order  is needed and help to pin down the overall position dependence of the deviation of fitting.

A correction vector $\vec {V_{ij}}$ of each spot, which shifts the measured values to the expected mean values, is employed:
\begin{equation}
\vec {V_{ij}} = (X_{expected,ij}-X_{measured,ij})\vec {e_{x}} +(Y_{expected,ij}-Y_{measured,ij})\vec {e_{y}},
\label{vector-ij}
\end{equation}
where $X_{expected,ij}$, $Y_{expected,ij} $ are the expected X value and Y value, $X_{measured,ij} $, $Y_{measured,ij} $ the measured  X value and  Y value,  and $\vec {e_{x}}$ and $\vec {e_{y}}$ the unit vectors in X- and Y-direction. The resulting correction matrix $\vec {V_{ij}}$ shifts each measured mean value to its expected position on the mask. 
A 2D vector field figure as shown in  Fig.~\ref{fig7} displays the correction vector $\vec {V_{ij}}$ of each spot shifts on the mask with the angle and magnitude (line length) shown. For better visibility, the  lengths (magnitudes)  of the vectors are enlarged by a factor of 5. The edge of the active area of the MCP is marked by the dashed blue circle. As can be seen in the Fig.~\ref{fig7}, the deviation amplitude  of the third order fitting is much smaller at each point than the first order fitting except few points near the edge of the delay-lines, which verifies that higher order correction is necessary and results in a better accuracy. Besides, a curl in the direction of the vectors is not observed in the first order fitting vector field map, the negligence of a rotation in the XY-plane seems to be reasonable. From the first and higher  order correction map in Fig.~\ref{fig7},  larger deviations appear near the edge of the delay-lines.
\section{Experimental test of the electrostatic detector}
\label{experiment}
Conversion foil utilized for test in this paper is made of mylar (2 $\mu m$) coated with aluminium ($\sim$ 0.1 $\mu m$). The distance of the accelerating grid to the foil  is kept at 8 mm. The grid wire spacing are all set as 1 mm. 
\subsection{Experimental setup}
\begin{figure}[!bh]
\centering
	\includegraphics[angle=0,width=0.499\textwidth]{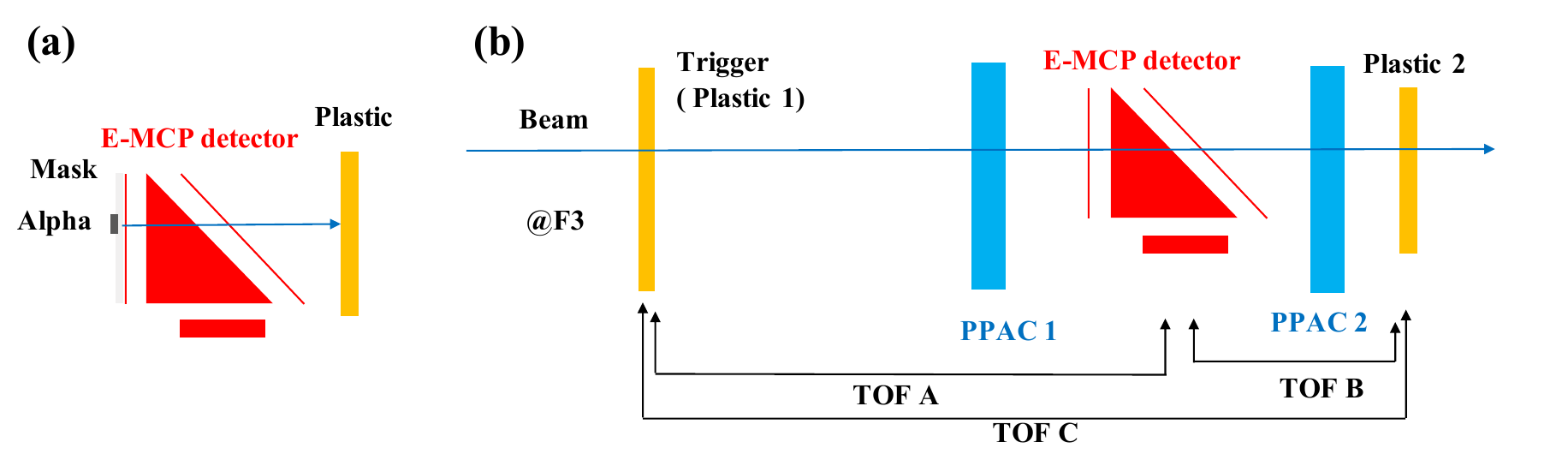}
	\caption{(Colour online)  Schematic view of detector arrangement of the offline (a) experiment with $\alpha$ source  and online (b) experiment with heavy ion beams. 
		\label{fig8}}
	\end{figure}
\subsubsection{Offline setup}
\label{offline setup}
The offline test setup is shown in Fig.\ref{fig8}(a). A mask with many holes dug in diameter of 1 mm or 0.5 mm are designed to collimate the $\alpha$ source  to test the position resolution of the electrostatic MCP (E-MCP) detector. Three $^{241}$Am $\alpha$ sources are pasted on the mask which is fixed onto the foil-plate by screws. A plastic scintillation counter  is installed right after the E-MCP detector to accept the passing though alpha. The plastic scintillation counter consists of a plastic scintillator and two PMTs (model number H2431-50~\cite{PMT}) which are connected to both ends of the scintillator. Coincidence signal of the left and right signals from the 2 PMTs of the plastic scintillator  is utilized as trigger for the CAMAC Data Acquisition (DAQ) system.
\subsubsection{Online setup}
\label{online setup}

An online experiment was carried out at the secondary beam line, SB2 course 
 in Heavy Ion Medical Accelerator in Chiba (HIMAC)
 at National Institute of Radiological Science  (NIRS)~\cite{HIMAC,HIMACSB2}, Japan.
A primary beam of  $^{84}$Kr$^{36+}$ at an energy of 200 MeV/nulceon was used to test the performance of the E-MCP detector.
A schematic view of the experimental setup is shown in right panel of Fig.~\ref{fig8}. 
The setup consists of two delay-line  parallel plate avalanche chambers (PPACs)~\cite{PPAC}, one electrostatic MCP (E-MCP) detector, two plastic scintillator counters.
The position of each ion are determined and tracked by two gas filled detectors PPACs (PPAC1 and PPAC2) with a size of 100 mm $\times$ 100 mm placed in between the two plastic scintillation counters. 
Each of the two plastic scintillation counters consists of one scintillator and two PMTs (model number H2431-50~\cite{PMT}) at both ends of the scintillator.
The E-MCP detector was placed between PPAC1 and PPAC2. 
The position of each ion when passing through the foil  is reconstructed by the two PPACs.
\subsection{Experimental results of the electrostatic MCP detector}
\label{electrostatic MCP detector}
\subsubsection{Detection efficiency}
\label{efficiency}
\begin{figure}[!hbt]
\centering
	\includegraphics[angle=0,width=9.0 cm]{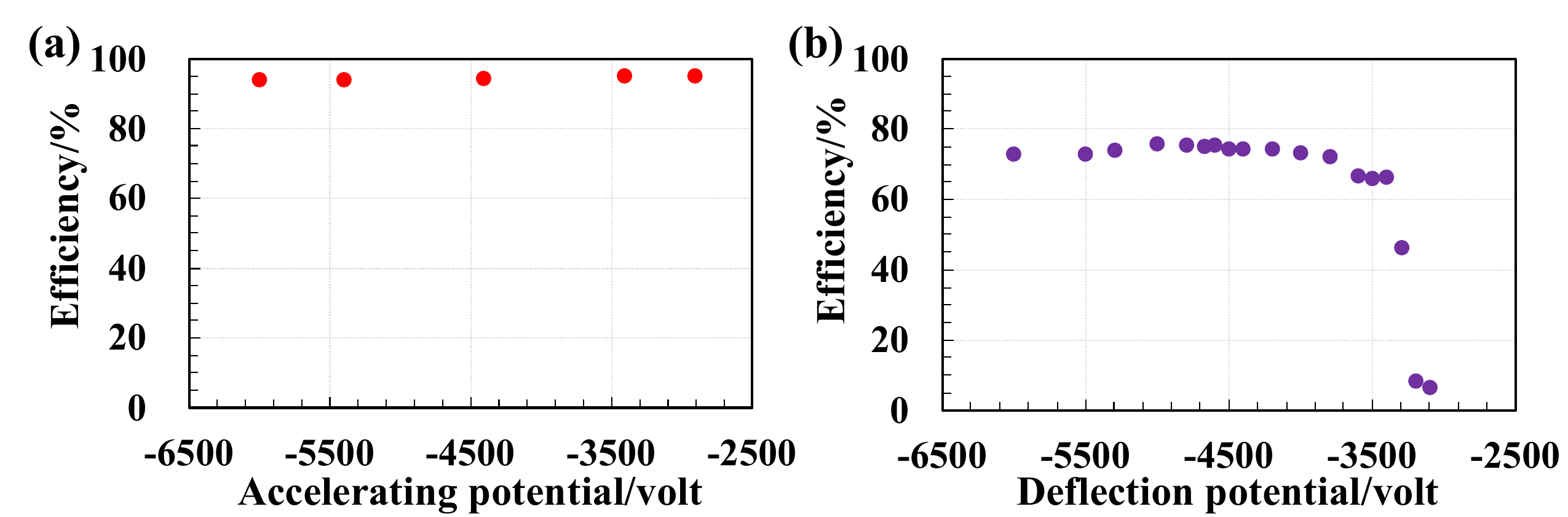}
	\caption{(a) The global detection efficiency of the electrostatic detector as a function of the deflection potential for the ion of $^{84}Kr^{36+}$. The ratio of deflection potential and accelerating potential is kept at $\sim$ 0.79. (b) The detection efficiency as a function of the deflection potential for $\alpha$ particles from $^{241}$Am source. The accelerating potential is kept at -6000 V and  the deflection potential is varied.
	\label{fig9}}
\end{figure}

\begin{figure}[!hbt]\centering
	\includegraphics[angle=0,width=0.49\textwidth]{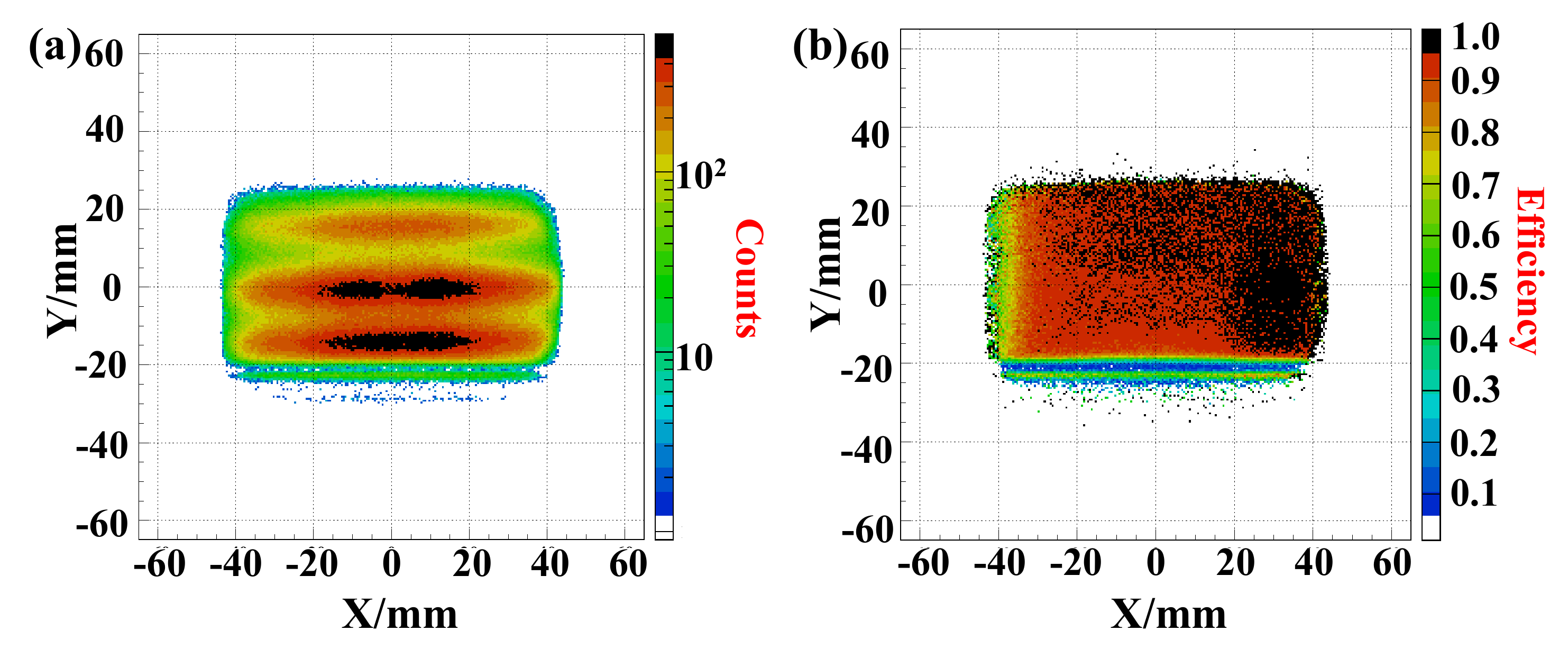}
	\caption{(Colour online) (a) shows the measured position distribution of beam by the electrostatic detector. (b) indicates the local detection efficiency distribution of the electrostatic detector.
	\label{fig10}}
\end{figure}

Efficiency is one of the most basic characteristics reflecting the probability for a certain species of particle penetrating through the conversion foil to be recorded by the DLD. Efficiency taken in the offline and online tests are corresponding  to the overall events measured by the E-MCP detector over the gated trigger events. The detection efficiency of the E-MCP detector is checked by  heavy ion beams and also $\alpha$ particles from $^{241}$Am source and the results are shown in Fig.~\ref{fig9} and Fig.~\ref{fig10}. Fig.~\ref{fig9}(a) demonstrates a steady efficiency of $\sim$ 95$\%$ as a function of the the deflection potential for heavy ions of $^{84}$Kr$^{36+}$ by arranging the ratio of deflection potential and accelerating potential to be kept at $\sim$  0.79 and varying the deflection potential (HV of the outer mirror grid). In Fig.~\ref{fig9}(b), the detection efficiency as a function of the deflection potential for $\alpha$ particles from $^{241}$Am source is demonstrated. The accelerating potential is kept at -6000 V and the deflection potential is varied during the test by changing the outer mirror grid HV. As can be seen from Fig.~\ref{fig9}(b), the detection is significantly dropping due to the  ratio of deflection potential and accelerating potential close to 0.5, which makes the mirror transparency to SEs as indicated by the Eq.~(\ref{transparency}). At the effective area of the MCP, a steady efficiency of $\sim$ 75$\%$ for the detection of $\alpha$ particles from $^{241}$Am source is obtained. The global efficiency is calculated using all the detected events on the whole MCP detector over the incident events, assuming that the detection efficiency has no dependence on the detection position of the detector. 

Local detection efficiency of the E-MCP detector is checked with 3  accumulated defocus beams which covers an area of about 90 mm $\times$ 50 mm. Fig.~\ref{fig10}(a) illustrates the 2-dimentional (2D) histogram of beam positions measured by the E-MCP detector.
The local detection efficiency of the E-MCP detector is  displayed by a 2D histogram with the X- and Y-coordinate bin sizes of 0.5 mm as shown in Fig.~\ref{fig10}(b). The efficiency is the measured event number by the E-MCP detector over the gated trigger events in bins with sizes of 0.5 mm for both X- and Y-coordinate. The  results in Fig.~\ref{fig10}(b) demonstrate small deviations in the the local efficiency. Especially on the left edge, there is a significant lack in the efficiency. The reason for this effect could be an inhomogeneous gain factor of the MCP stack and it can be solved by providing higher HV between the MCP stack to increasing the gain.
\subsubsection{Timing resolution}
\label{timing-resolution}
\begin{figure}[!hbt]\centering
	\includegraphics[angle=0,width=0.3\textwidth]{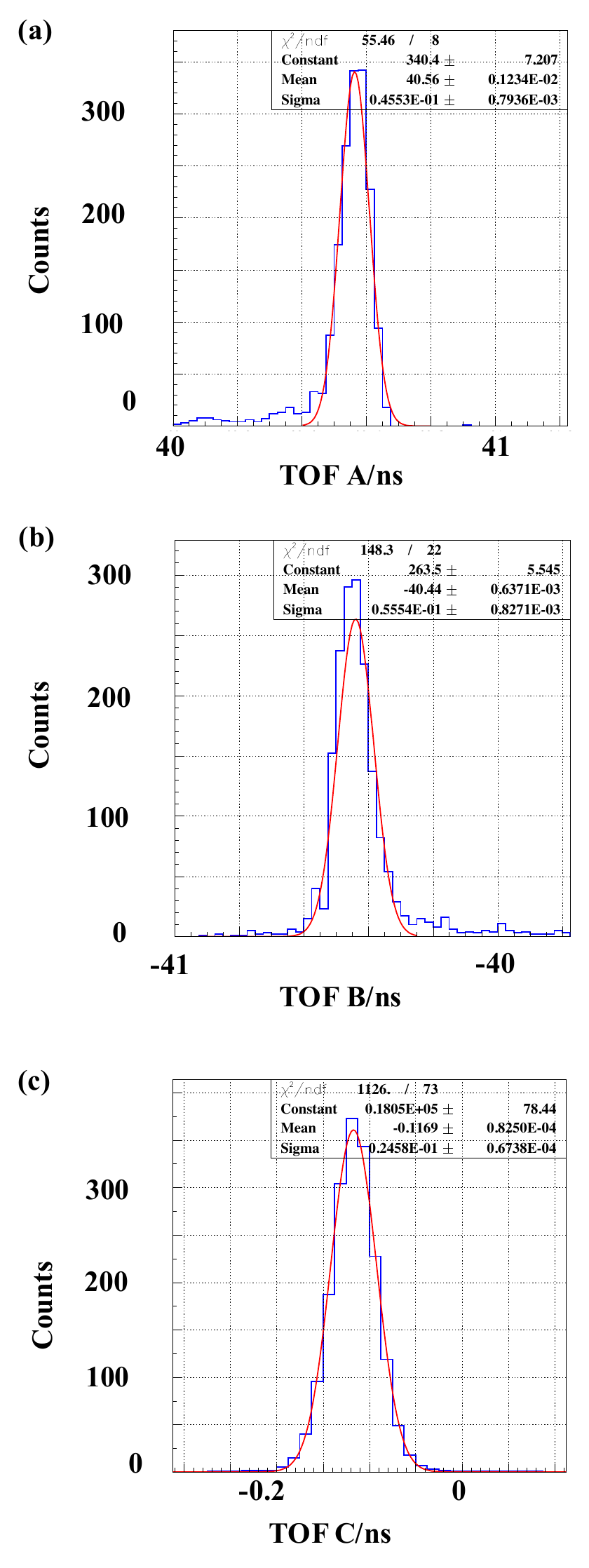}
	\caption{(Colour online) (a), (b) and (c) are spectra of three TOF sets for the intrinsic timing resolution deduction of the E-MCP detector. The mean parameter at right part of each panel means the mean value of the gaussian fitting of the TOF spectrum. The sigma value is the uncertainty of each TOF set, including the intrinsic resolutions of two timing detectors. 
	\label{fig11}}
\end{figure}

Normally, for one time-of-flight (TOF) spectrum, the distribution contains the information of two timing detectors, which is characterized by the uncertainty $\sigma$ or FWHM of the distribution. To determine the intrinsic timing resolution of the E-MCP detector, another  timing detector have to be applied as shown in Fig.~\ref{fig8}.
Assuming that the timing distribution of each timing detector follows a Gaussian distribution, the timing resolution of an individual detector can be determined by using three sets of TOF between three timing detectors: $TOF(A)$, $TOF(B)$, and $TOF(C)$. The 3 sets of TOFs ($TOF(A)$, $TOF(B)$, and $TOF(C)$) are demonstrated in Fig.~\ref{fig11}.
The resolutions of three TOFs describes as: 
\begin{equation}
(\sigma_{TOF(A)})^2= (\sigma_{pla1})^2 + (\sigma_{MCP})^2,
\label{intrinsic-timing-equation1}
\end{equation}
\begin{equation}
(\sigma_{TOF(B)})^2= (\sigma_{MCP})^2 + (\sigma_{pla2})^2,
\label{intrinsic-timing-equation2}
\end{equation}
\begin{equation}
(\sigma_{TOF(C)})^2= (\sigma_{pla2})^2 + (\sigma_{pla1})^2,
\label{intrinsic-timing-equation3}
\end{equation}
where $\sigma_{pla1}$, $\sigma_{MCP}$, and $\sigma_{pla2}$ are intrinsic timing resolution of plastic scintillator 1, the  E-MCP detector, and plastic scintillator 2, respectively. The $\sigma(TOF(A))$ (or $\sigma(A)$), $\sigma(TOF(B))$ (or $\sigma(B)$),  $\sigma(TOF(C))$ (or $\sigma(C)$) are the fitting 'sigma' parameters of the 3 TOF sets.  From these relationships, we can deduce the intrinsic timing resolution of E-MCP detector by solving the above equations:
\begin{equation}\small
\sigma_{MCP} = \sqrt{(\sigma^{2}(TOF(A)) + \sigma^{2}(TOF(B)) -  \sigma^{2}(TOF(C)))/2},
\label{intrinsic-timing-resolution-solution}
\end{equation}
where the $\sigma_{MCP}$ indicates the intrinsic timing resolution of the E-MCP detector.
The uncertainty $\delta(\sigma_{mcp})$ of the timing resolution $\sigma_{MCP}$ from the measurement can be expressed as:
\begin{equation}\scriptsize
\delta(\sigma_{mcp})=\sqrt{(\delta^{2}(\sigma(A))\sigma^{2}(A)+\delta^{2}(\sigma(B))\sigma^{2}(B)+\delta^{2}(\sigma(C))(\sigma^{2}(C))/(4\sigma^{2}_{MCP})},
\label{intrinsic-timing-resolution-uncertaity}
\end{equation}
where $\delta(\sigma(A))$, $\delta(\sigma(B))$, $\delta(\sigma(C))$ are the uncertainties of the fitting ``sigma'' parameters, $\sigma(TOF(A))$ (or $\sigma(A)$), 
$\sigma(TOF(B))$ (or $\sigma(B)$),  $\sigma(TOF(C))$ (or $\sigma(C)$), and the $\delta(\sigma_{mcp})$ indicates the uncertainty of the deduced value of the intrinsic timing resolution $\sigma_{MCP}$ of the E-MCP detector. One example of the spectra of three TOF sets for the intrinsic timing resolution deduction of the timing detectors  is demonstrated in Fig.~\ref{fig11}. The ``sigma'' parameter and the fitting uncertainty of it correspond to the $\sigma$ and $\delta(\sigma)$ values in Eq.~(\ref{intrinsic-timing-resolution-solution}) and  Eq.~(\ref{intrinsic-timing-resolution-uncertaity}) respectively. 

\begin{figure}[!hbt]\centering
	\includegraphics[angle=0,width=0.45\textwidth]{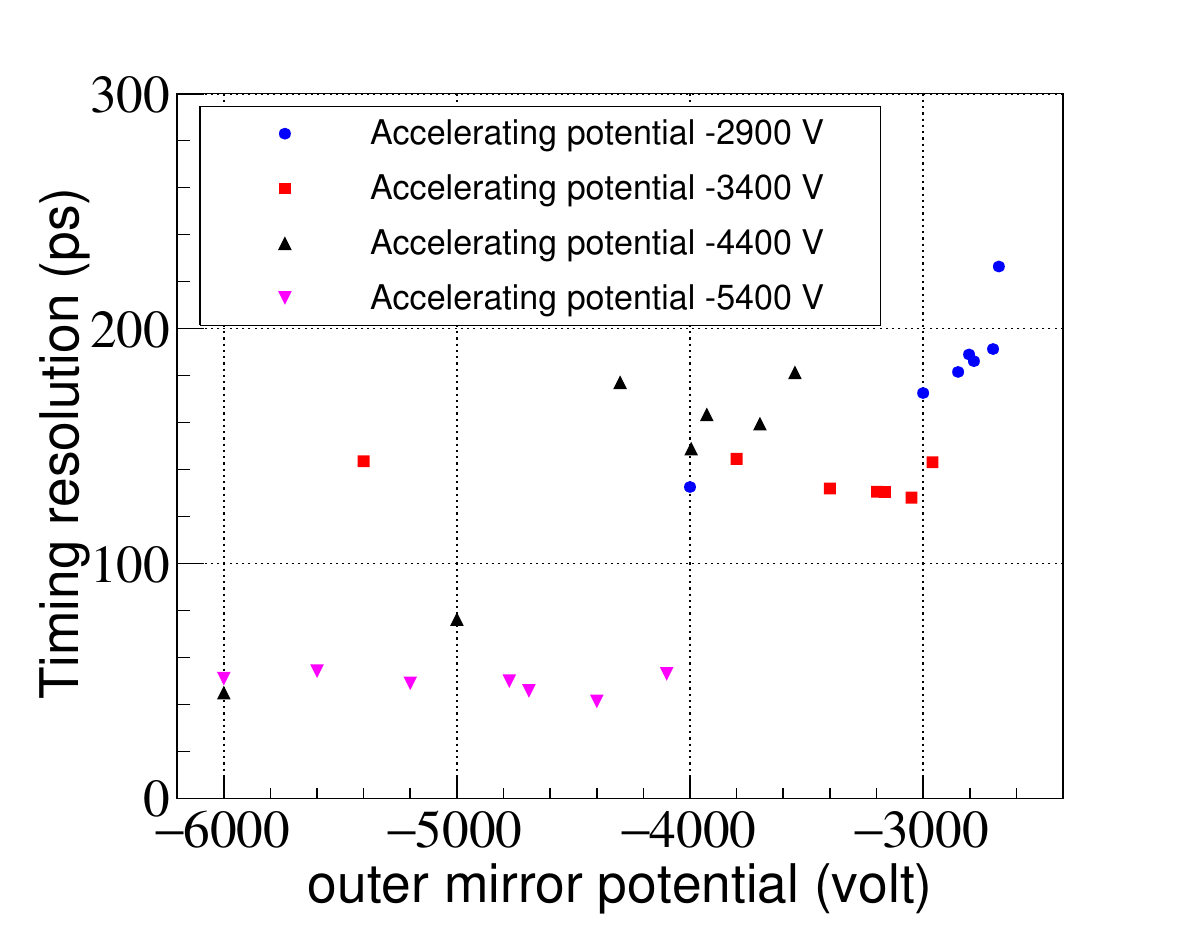}
	\caption{(Colour online) 
	 Timing resolution of the E-MCP detector  for $^{84}$Kr$^{36+}$ ions as a function of the outer mirror potential by keeping the accelerating potential at several certain values indicated by different colors as shown in the legend.  The error bars are smaller than the symbol size and thus not visible.
	\label{fig11-2}}
\end{figure}
\begin{figure}[!hbt]\centering
	\includegraphics[angle=0,width=0.45\textwidth] {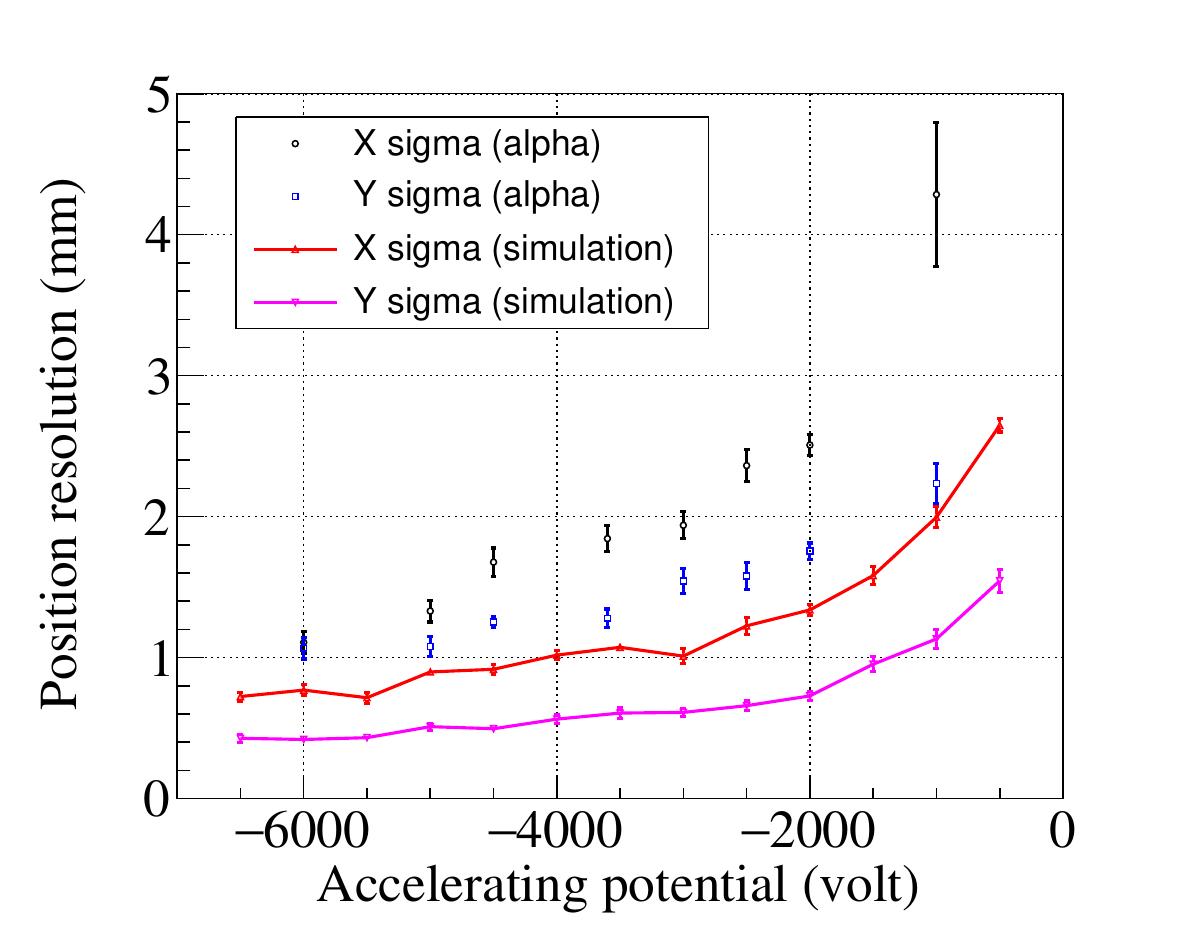}
	\caption{(Colour online) Position resolution comparison of offline results (mylar foil in ``electron mode'') to simulation results as a function of the accelerating potential by keeping the ratio of accelerating potential and the deflection potential at $\sim$ 0.778.
	\label{fig13}}
\end{figure}



\begin{figure}[!hbt]\centering
	\includegraphics[angle=0,width=0.48\textwidth]{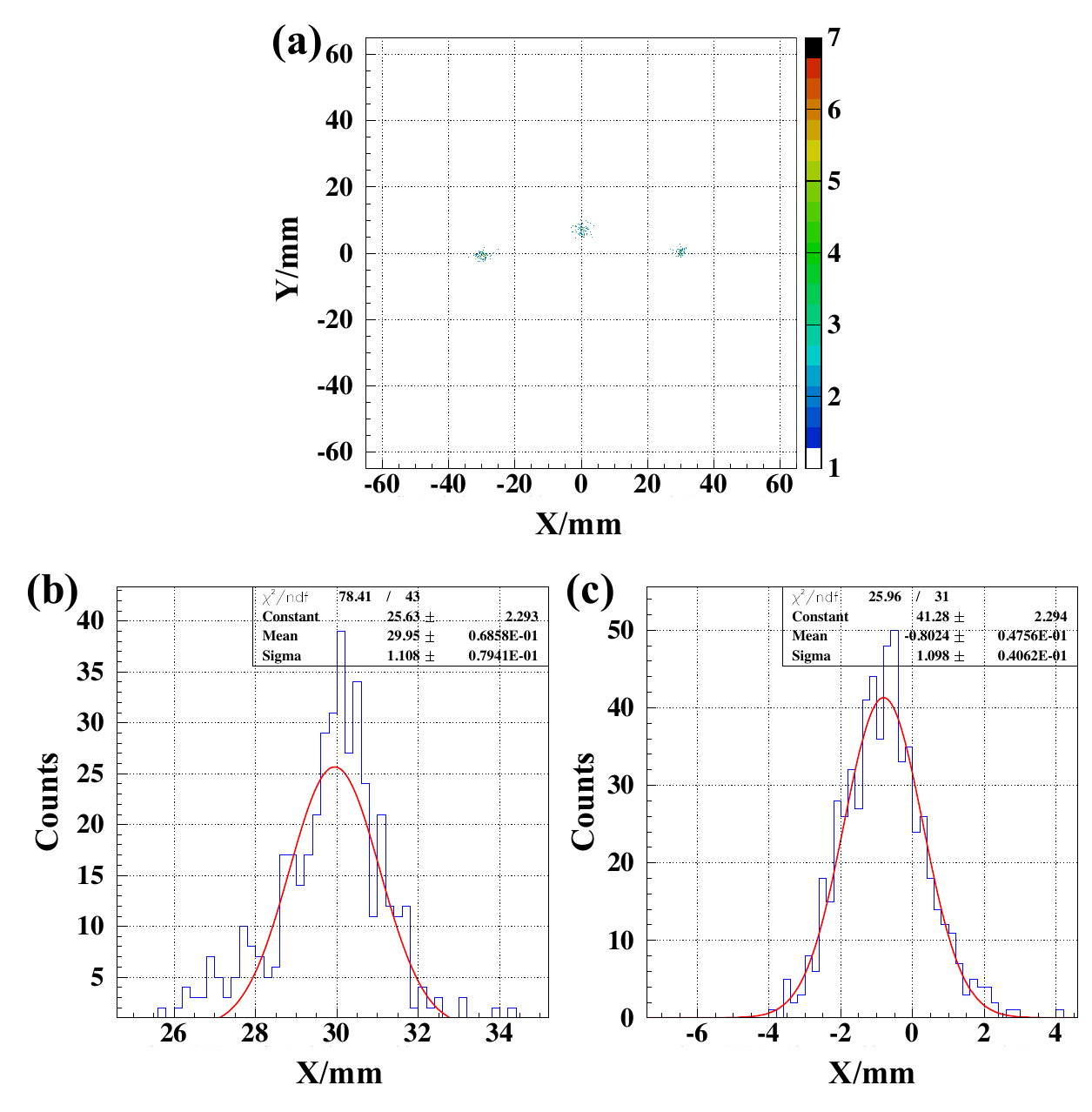}
	\caption{(Colour online) (a) is a imaging of collimated $\alpha$ particles from 3 holes on a mask in front of the foil.  (b) and (c) are  X and Y-coordinate projections of the  imaging of one hole.  The gaussian fitting parameter `sigma' of the peak  is utilized to characterize the resolutions (X: 1.108 mm, Y: 1.098 mm). The deviations between of the imaging points on the MCP detector and their corresponding physical positions on the mask are smaller than 1$\sigma$ uncertainty (the resolution) of the measurements.
	\label{fig12}}
\end{figure}
The intrinsic timing  resolution  as a function of the outer mirror potential by changing the accelerating potential within four different setting values is  shown  in Fig.~\ref{fig11-2}. 
The best obtained timing resolution is $43 \pm 2$ ps in $\sigma$.
\subsubsection{Position resolution}
\label{position-resolution}
The position resolution for the E-MCP detector in the offline test  is checked by using collimated (hole sizes smaller than 0.5 mm in diameter) $\alpha$ from $^{241}$Am  in front of the conversion foil.
The black and blue data points in  Fig.~\ref{fig13} shows the X and Y direction position resolution as a  function of accelerating potential respectively. The  ratio of accelerating potential and the deflection potential is kept at $\sim$ 0.778. 
The simulation result by SIMION~\cite{SIMION} with the same setting of the HV supplies for the detector are indicated in solid line in Fig.~\ref{fig13}.  The red  corresponds to the X direction resolution and pink represents the Y direction resolution. The differences of the experimental results and simulation results are mainly from the inhomogeneous electrostatic field,  timing walk in electronics,  spread of electrons in the MCP and the  initial condition difference of the real condition and simulation.
It is obvious that as the increasing of the HV of accelerating potential and the deflection potential, the resolutions of X and Y direction get improved, which is consistent with the simulation results as shown in Fig.~\ref{fig13}.

In one offline test run, we set 3 $\alpha$ sources ($^{241}$Am) with nearly same intensity of 4 MBq at the hole places of (-30 mm, 0 mm), (0 mm, 8 mm), and (30 mm, 0 mm) on the mask with hole size of less than 0.5 mm  in diameter. The imaging of collimated $\alpha$ particles from the holes are shown in Fig.~\ref{fig12}(a). The projections on the X- and Y-coordinate of the imaging of one hole  are shown in Fig.~\ref{fig12}(b) and Fig.~\ref{fig12}(c), respectively. The uncertainty of the gaussian fit for position distribution on the X- and Y-coordinate are 1.108 $\pm$ 0.079 mm and 1.098 $\pm$ 0.041 mm respectively. In this setting, the accelerating potential of  -6000 V and deflection potential of -4668 V were supplied.

\begin{figure}[!hbt]
\centering
	\includegraphics[angle=0,width=0.48\textwidth]{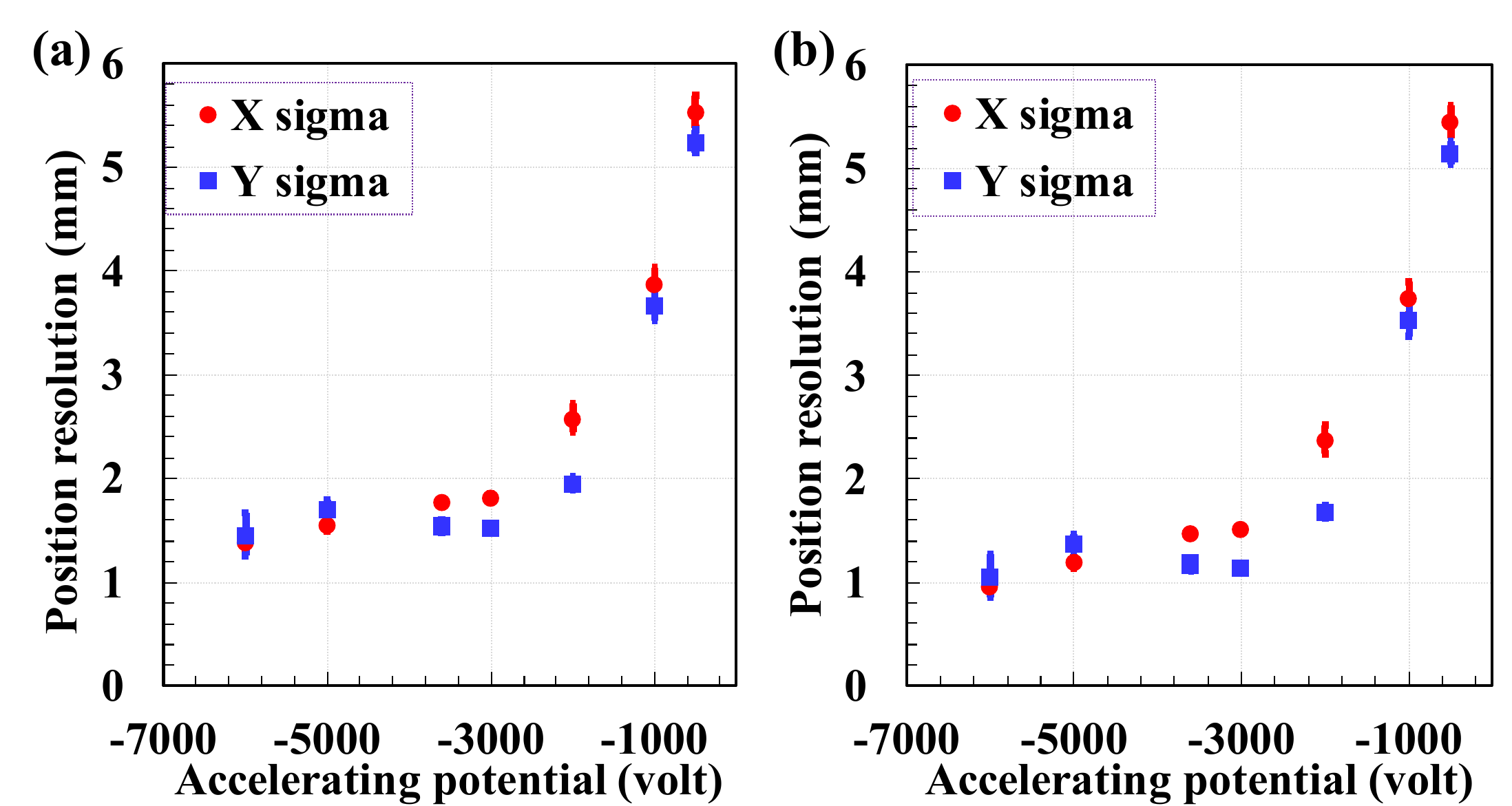}
	\caption{(Colour online) (a) Uncertainty of position measurement difference  of  the PPACs and the E-MCP detector as a function of accelerating potential. (b) Uncertainty of position measurement difference  subtracted with the resolution of the PPACs system (assuming a resolution of 1 mm for 2 dimensions)  as a function of accelerating potential.
	\label{fig14}}
\end{figure}

In the online experiment, the position on the  the foil  when  heavy ions passing though are measured by the E-MCP detector and also reconstructed by the PPACs. Uncertainty of position measurement difference  of  the PPACs and the E-MCP detector as a function of accelerating potential is shown in Fig.~\ref{fig14}(a). As the uncertainty of the position difference contains the intrinsic resolution of the E-MCP detector and the PPACs tracking system. To demonstrate the intrinsic resolution of the E-MCP detector, position resolutions of 1 mm for 2 dimensions of the PPACs tracking system are assumed to be subtracted from the uncertainty of the position difference. The derived intrinsic resolution of the E-MCP detector system, after subtraction of  resolution of PPACs tracking system, as a function of the accelerating potential is shown in Fig.~\ref{fig14}(b). The trend of  position resolution as a function of the accelerating potential is consistent with those of offline test and simulation shown in Fig.~\ref{fig13}. 
\begin{figure}[!hbt]
\centering
	\includegraphics[angle=0,width=0.4\textwidth]{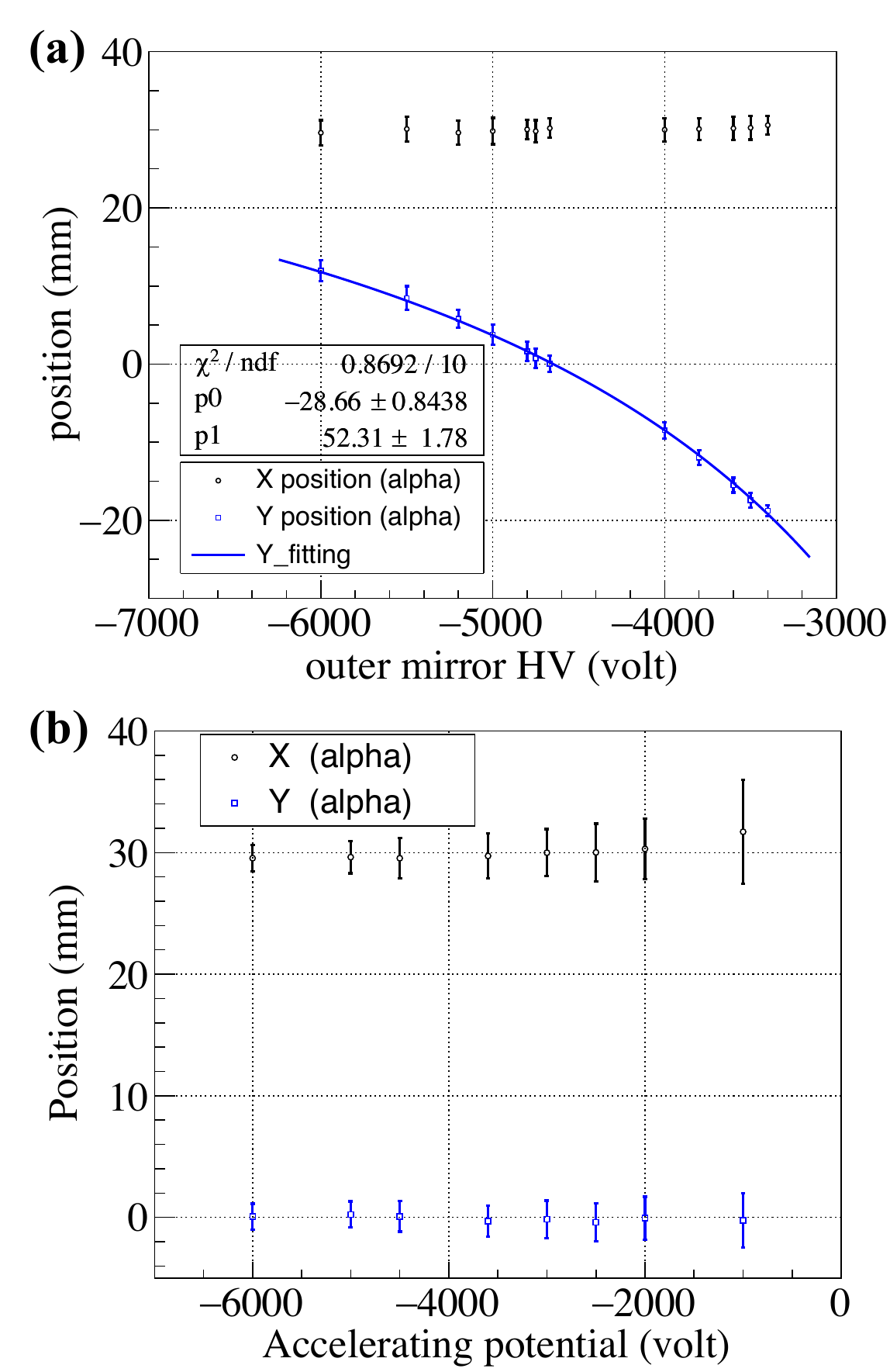}
	\caption{(Colour online) (a) shows the X , Y position of the SEs deflected onto MCP by the electrostatic detector deduced from collimated $\alpha$ source  at position (30 mm, 0 mm) on mask with a hole size of $\sim$ 0.5 mm in diameter, as a function of outer mirror potential while keeping the accelerating grid HV of -6000 V. (b) displays the X , Y position of the imaging of the same collimated hole as a function of accelerating potential by keeping the ratio of potential of the deflection potential to accelerating potential as a constant.
		\label{fig15}}
\end{figure}
\subsection{Trajectory confirmation of secondary electrons by experimental data}
\label{trajectory-section}
The trajectory of the SEs inside the electrostatic detector has been calculated theoretically as shown in section~\ref{motion}. The equation are demonstrated in Eq.~(\ref{eq_motion}) indicating that the position of Y direction of the imaging as functions of the accelerating potential and deflection potential. The X, Y direction position of the imaging by varying outer mirror HV and keeping the accelerating potential at -6000 V  are drawn at Fig.~\ref{fig15}(a) and the Y direction motion of the imaging is fit with $y_{exp} ={{P{0}}}/{2}\cdot\frac{U_{acc}}{U_{mir}}+P1$ (from Eq.~(\ref{eq_motion})), where  the  parameters P0  represents the distance between mirror grids of 28 mm in design and P1 indicates the shift of the SEs in the mirror along beam direction. Trajectory and mirror distance can be well reproduced as shown in Fig.~\ref{fig15}(a) with the P0 value of 28.66 $\pm$ 0.84 mm and P1 value of 52.31 $\pm$ 1.78 mm, which are very close to the designed value of 28 mm and 52.672 mm (47.672 mm designed value and  5 mm shift in analysis code), respectively. Figure~\ref{fig15}(b) shows the imaging position of X- and Y-coordinate of the collimated hole (30 mm, 0 mm) on DLD. As the ratio of the deflection potential to the accelerating potential is kept as a constant of 0.778 and the X, Y direction of the imaging will be the same for all the setting by varying the deflection and accelerating potential at the same time.

\section{Summary}
\label{summary}
A prototype foil-MCP detector, which possesses specifications of low energy loss and energy straggling by detection of secondary electrons (SEs) induced from conversion foil by heavy ions, a large effective area ($\sim$ 110 mm $\times$ 45 mm) to cover a large beam size, good timing and position resolution at the same time, has been developed for mass measurements and beam monitoring at the Rare-RI Ring. The performance of timing and position measurements for the detector has been studied and characterized systematically. To characterize and optimize the timing and position resolution of the detector, resolution dependence of high voltage supplies has been studied experimentally and by simulation. Especially, an isochronous condition has been studied for this type of detector. Experiments aimed at studying the performance of the detector were conducted at HIMAC (Heavy Ion Medical Accelerator in Chiba) with heavy ion beam and with $\alpha$ source  of $^{241}$Am. The performance of the detector have been optimized and the best achieved timing resolution is better than 50 ps (in $\sigma$) and position resolution $\sim$ 1 mm (in $\sigma$) for 2-dimensions respectively, for which the detection efficiency is $\sim$ 95$\%$  for heavy ion beam. The overall efficiency is $\sim$ 75$\%$  for $\alpha$ particles from $^{241}$Am source.

This type of detector is a versatile instrument which can be used on the beam-line for two dimensional position measurement to reconstruct beam trajectory, beam-line momentum measurements for velocity reconstruction. Meanwhile, it can be used for position monitoring and revolution time measurement turn by turn inside the storage ring, R3. Low energy loss and small angular straggling of the detected incident ions are indispensable for reconstruction of the velocity for in-ring mass deduction or momentum measurement with good accuracy and high precision for B$\rho$-TOF mass measurements. High-resolution TOF measurement in ring is not only significant for identification of nuclei with high resolving power but also can be used for mass measurements directly. The fast low-energy-loss position-sensitive timing detector developed within this paper will be utilized for in-ring and on beam-line to realize a higher performance mass measurements simultaneously employing the Rare-RI Ring in conjunction with the high-resolution beam-lines BigRIPS-OEDO-SHARAQ by the two complementary TOF Methods: Isochronous Storage Ring Mass Spectrometer (IMS) and Magnetic-rigidity-time-of-flight (B$\rho$-TOF).

\section*{Acknowledgments}
We would like to acknowledge all other members of the Rare-RI Ring collaboration, and all technicians of the HIMAC accelerators, who made the experiments possible to test the detector and gave valuable suggestions in discussion. 

\end{document}